\pdfoutput=1

\documentclass[a4paper,10pt]{scrartcl}
\usepackage{amsthm}
\usepackage[utf8x]{inputenc}
\usepackage{ISASPackages/ISASMatheMacros}
\usepackage{mathtools}
\usepackage{graphicx}
\usepackage[squaren,thinspace]{SIunits}
\usepackage{hyperref}
\usepackage{subfigure}
\newcommand{\erw}[2][]{
     \ensuremath{\mathbb{E}_{#1} \! \left\lbrace #2 \right\rbrace}%
}

\renewcommand{\Pr}{\mathrm{Pr}}
\renewcommand{\Cov}{\mathrm{Cov}}

\newtheorem{lemma}{Lemma}
\newtheorem{thm}{Theorem}
\newtheorem{definition}{Definition}

\title{Chance-constrained Model Predictive Control for Multi-Agent Systems}
\author{Daniel Lyons% 
	\thanks{D. Lyons is with the Intelligent Sensor-Actuator-Systems Laboratory (ISAS), Karlsruhe Institute of Technology (KIT), Karlsruhe, Germany.
        {\tt\small lyons@kit.edu}.}%
, Jan-P. Calliess% 
	\thanks{J. Calliess is with the Robotics Research Group, University of Oxford, Oxford, OX1 3PJ, UK.
	{\tt\small jan@robots.ox.ac.uk}.}%
, and Uwe D. Hanebeck%
\thanks{U. D. Hanebeck is with the Intelligent Sensor-Actuator-Systems Laboratory (ISAS), Karlsruhe Institute of Technology (KIT), Karlsruhe, Germany.
        {\tt\small uwe.hanebeck@ieee.org}.}%
}

\begin{document}

\maketitle

\begin{abstract}

We consider stochastic model predictive control of a multi-agent systems with constraints on the probabilities of inter-agent collisions. 
% If the states of the agents are affected by stochastic disturbances, a probabilistic modeling of inter-agent collision avoidance constraints will generally be safer than constraints that ignore these disturbances and less conservative than robust constraints that will try to account for the strongest possible disturbances. 
% There are two main challenges when considering stochastic constraints for collision avoidance. 
% One challenge is that the collision probabilities cannot be determined efficiently since multivariate integrals have to be computed and the other is that the feasible region for the stochastic optimization problem is not necessarily convex anymore.  
We first study a sample-based approximation of the collision probabilities and use this approximation to formulate constraints for the stochastic control problem. 
This approximation will converge as the number of samples goes to infinity, however, the complexity of the resulting control problem is so high that this approach proves unsuitable for control under real-time requirements. 
To alleviate the computational burden we propose a second approach that uses probabilistic bounds to determine regions with increased probability of presence for each agent and formulate constraints for the control problem that guarantee that these regions will not overlap. 
We prove that the resulting problem is conservative for the original problem with probabilistic constraints, ie. every control strategy that is feasible under our new constraints will automatically be feasible for the original problem.
Furthermore we show in simulations in a UAV path planning scenario that our proposed approach grants significantly better run-time performance compared to a controller with the sample-based approximation with only a small degree of sub-optimality resulting from the conservativeness of our new approach. 
\end{abstract}

\section{Introduction}
There are many applications, in which the deployment of multiple robots or UAVs is advantageous for the completion of the mission compared to just a single robot or UAV.
For example, the search for a target with multiple UAVs \cite{Hoffmann2010} or localization of an odor source with multiple robots \cite{Hayes2002} have been studied in the past.
% Further applications that have been studied are cooperative load transportation with multiple UAVs \cite{Kondak2009} or the cooperative inspection of airplane turbines with miniature robots \cite{Correll2006}.
In all of these applications, path planning, obstacle avoidance, and collision avoidance with other robots also play an important role, since without path planning and collision avoidance, the successful completion of the task at hand and the physical intactness of the robots or UAVs cannot be warranted.
We will from now on not further distinguish between a robot, a UAV, or a ground vehicle and will therefore call the entity making path planning decisions an ``agent''.
% Depending on the complexity of the agents, path planning and collision avoidance strategies can consist of very simple heuristics or of elaborate plans that take considerations like cost optimality or guarantees for reaching the target into account.
% Especially for multi-agent systems consisting of expensive and complex agents, equipped with limited energy resources but with extended computational power, optimality and reliability are of increased importance.

% The quality of obstacle avoidance and collision avoidance with other agents depends also on the quality of an agent's localization of itself and the other agents (and how good its own motion model is).
Noisy sensor measurements for localization and imprecise models of the motion dynamics can lead to uncertain estimates of the pose of the agents.
When ignored by the path planning algorithm, these uncertain estimates can cause a failure of the mission through collisions with obstacles or other agents and so it is crucial to account for them while planning. 
% As a consequence
% An established way to model uncertainties in position estimates is through a probabilistic approach, where the positions of the agents are modeled as random variables.
% In planning, uncertainties are often accounted for through probabilistic chance constraints on the failure of the mission \cite{Birge1997,Schwarm1999}.
% Another factor that increases safety and optimality of planning is planning model predictively over a receding time horizon. 
% Here the agents do not only aim to find actions that bring them closer to their target in a myopic way, but try to find a plan that is optimal looking forward in time for several time instances \cite{Bertsekas1995}. 
%, and hence the plans become more sustainable and safer \cite{Bertsekas1995}. 
% Stochastic model predictive control has the drawback that the planning problem can become very complex and consequently it is vital to pay attention to real-time capabilities of planning. 

If a stochastic description of the agents' states is considered, constraints on the behavior of the agents should also be treated in a probabilistic manner. 
Probabilistic chance constraints are constraints on the probability that the agents are in states that could cause a failure of the mission, i.e., collide with obstacles or with each other \cite{Schwarm1999,Birge1997,Prekopa1995}. 
In contrast to robust control, where one tries to find control actions that are optimal under bounded disturbances, chance constraints allow to model problems with disturbances with arbitrary probability distributions. 
In order to make these probabilistic constraints computationally feasible, the distributions over the agents' states have to be either very simple (e.g. Gaussian) with simple constraints (linear or convex) or the distributions from which the constraints are constructed have to approximated. 

Another factor that increases safety and optimality of planning is planning model predictively into the future. 
In model predictive control the controller uses a model of the agents dynamics to extrapolate the agents' states under candidate control sequences several time steps into the future and then chooses the control input that is not only optimal for the current states but also for the extrapolated states. 
Hence, model predictive control (MPC) enables the agents to plan pro-actively to avoid obstacles or other agents, since conflicts or possible collisions are detected earlier and the agents can react more quickly and more efficiently to avoid them. %, and hence the plans become more sustainable and safer \cite{Bertsekas1995}. 
% Here the agents do not only aim to find actions that bring them closer to their target in a myopic way, but try to find a plan that is optimal looking forward in time for several time instances \cite{Bertsekas1995}. 
However, stochastic MPC has the drawback that the planning problem can become very complex and consequently it is vital to pay attention to real-time capabilities of planning. 
% As soon as the probability distributions involved become more complex and possibly multivariate or multi-modal or the constraints are no longer convex, there is generally no closed form formulation and the chance constraints have to be approximated.
% The challenge, however, with MILPs is, that for programs with many integer variables runtime can increase exponentially and, hence, when formulating MILPs it is important to keep the number of integer variables as low as possible.

There are two main challenges when considering stochastic constraints for collision avoidance. One challenge is that the collision probabilities cannot be determined efficiently since multivariate integrals have to be computed and the other is that the feasible region for the stochastic optimization problem is not necessarily convex anymore.  

In this work, we propose two approaches to formulate stochastic model predictive path planning for multi-agent systems with chance constraints on the probability of inter-agent collisions as a mixed integer linear program (MILP). 
MILPs are a well-understood problems that can be solved very efficiently for moderate problem sizes and find many applications in robot planning, flight control, and receding horizon control (see for example \cite{Schouwenaars2001,Bellingham2002,Schouwenaars2004,Yilmaz2008}).
% These approaches are an extension of the work in \cite{Blackmore2010} to multi-agent systems. 
% The chance constraints refer to both the probability of an agent leaving the feasible region and to coupling constraints on the probability of a collision between agents (called inter-agent collisions from now on). 

% Both formulations differ only in their formulation of the inter-agent collision avoidance. 
We first study a sample-based approximation of the collision probabilities and use this approximation to formulate constraints for the stochastic control problem. This approximation will converge as the number of samples goes to infinity, however, the complexity of the resulting control problem is so high that this approach proves unsuitable for control under real-time requirements. 

To alleviate the computational burden we propose a second approach that uses probabilistic bounds to determine regions with increased probability of presence (RIPP) for each agent and formulate constraints for the control problem that guarantee that these regions will not overlap. 
We prove that the resulting problem is conservative for the original problem with probabilistic constraints, ie. every control strategy that is feasible under our new constraints will automatically be feasible for the original problem.
This is a very remarkable property of our novel constraints, since it guarantees feasibility for the original problem and we do not actually have to evaluate the complicated inter-agent collision probabilities. 

Since we employ a sample-based representation of the agents' uncertain positions and the probabilistic bounds we use for the RIPP regions hold for arbitrary uncertain states, we do not have to make any assumptions (such as being Gaussian) on the nature of the occurring noise or disturbances. 
Also in our approach we do not have to assume that the chance constraints are given through linear inequalities only and thus, are able to model the more complex and inherently non-convex coupling constraints on the states of the agents. 

We show in simulations in a UAV path planning scenario that our proposed RIPP approach grants significantly better run-time performance compared to a controller with the sample-based approximation of collision probabilities, with only a small degree of sub-optimality resulting from the conservativeness of our new approach. 
We also compare our probabilistic methods with robust control methods for multi-UAV collision avoidance and provide empirical evidence that our approach is better suited for stochastic settings, since it allows the user to precisely specify an upper bound on the probability of collisions for the UAVs. 

% The first formulation is a direct sample-based approximation of the chance constraints on inter-agent collisions. 
% It has the inconvenience that it produces an inflating number of binary constraints for the MILP that increases the runtime of the program significantly.

% To alleviate this computational burden, we propose a second formulation of the coupling constraints. % that conservatively bounds the probability of inter-agent collisions through constraints on the distance of the means of the position estimates of the agents.
% The Chebychev inequality determines the probability that a random variable with an arbitrary distribution deviates more than a predefined bound from its mean. 
% We determine the probability that the agent's true position is within a box around the expected value of its position estimate. 
% Then we set constraints on the distances of these boxes for each agent such that they do not intersect for different agents, allowing us to control the probability of an inter-agent collision and enforce the chance constraints on the probability of inter-agent collisions. 

To the best of our knowledge this is the first time a practical approach for the control of a multi-agent system with chance constraints on the probability of a collision of the agents is proposed. 
Existing work on chance constrained control deals either with planning for single agent systems or for systems without coupling chance constraints on joint states of different agents. 

% We analyze the complexity of both formulations in terms of the number of integer variables they introduce and show that the second formulation exhibits a much lower complexity, that is independent from the complexity of the sample representations of the state estimates of the agents.

% When considering planning in multi-agent systems one always has to make a decision between a central planning instance and decentralized control strategies. 
% Central planning has the advantage that the optimality of joint plans can be warranted, with the drawback of a higher communication overhead and forming a single point of failure. 
% Decentralized control strategies do not have these drawbacks but often times cannot guarantee optimality of their plans. 
% The formulations we give in this paper are at a priori central in nature, as the intended emphasis of this work is on the formulation of the chance constraint inter-agent collision avoidance as MILP constraints. 
% However, we also briefly discuss two means on how a possible decentralization would proceed. 
% One is a decentralization through Dantzig-Wolfe decomposition \cite{Bertsimas1997} and the other one through a priorization scheme of the agents. 

The paper is structured as follows. 
In Section~\ref{sec:gen_prob}, we formally define the general problem of model predictive control for a multi-agent system with chance constraints. 
In Sections~\ref{sec:sys_dynamics} and~\ref{sec:sa_milp}, we describe the system dynamics of the agents, the sample approximation of the state distributions, and outline how to formulate single agent planning as a MILP. 
The contribution of this paper lies in Sections~\ref{sec:col_agents} and~\ref{sec:col_av_app}, where we first show how to approximate the collision avoidance constraints directly with samples. Then, in Section~\ref{sec:col_av_app} we construct our novel approach employing regions of increased probability of agent presence, prove its conservativeness and compare its computational complexity to the full approximation of the collision avoidance constraints. 
In Section~\ref{sec:comp_ana} we give a theoretical comparison of the complexity of the optimization problems constructed in the previous two sections. 
In Section~\ref{sec:sim}, we compare our novel approach with a controller with the sample-based approximation in a UAV path planning scenario with non-Gaussian wind turbulence models. 
Section~\ref{sec:concl} concludes the paper and points to future work. 

\subsection{Related Work}
The body of work on interactions of agents in a multi-agent system is immense, so we focus here on recent results in chance constrained control of single-agent and multi-agent systems.

The most recent work on chance-constrained MPC for linear systems can roughly be classified into three parts: conservative approximations, control of systems with Gaussian disturbances and sample-based control. %, represented through linear matrix constraints \cite{Hessem2002,Ono2008,Blackmore2009}. 

The authors of \cite{Hessem2002} approximate chance constraints through \emph{conservative constraints}, that ellipsoids around the means of the state estimates are completely contained in the feasible region. 
Computational comparisons in \cite{Blackmore2009} and \cite{Ono2008} indicate however that, albeit being very fast, this approach introduces a high degree of sub-optimality through its conservativeness. 

For systems where the disturbance distributions are known to be \emph{Gaussian}, \cite{Ono2008, Blackmore2009} propose to employ an inequality from probability theory called Boole's inequality. 
% Another form of approximation is to conservatively replace the occurring multivariate Gaussian state distributions by axis-aligned Gaussians \cite{Ono2008, Blackmore2009}. 
Here, instead of enforcing that all state constraints are satisfied with a certain probability at the same time, each linear state constraint is considered separately and the probability of violating this constraint is also enforced separately. 
These 'separated' chance constraints can then be formulated through one-dimensional constraints on Gaussian cumulative distributions, making the resulting problem a convex optimization problem. 
The work \cite{Ono2010} extends these approaches to the control of multi-agent systems, however, without considering coupling constraints on the states of the agents that are necessary for modeling collision avoidance. 
Instead, the authors aim to find an optimal allocation of the overall probability of the failure of the mission among the agents in a decentralized manner. 

The work~\cite{Vitus2011} uses results from linear quadratic control with Gaussian disturbances and measurement noise to derive closed-loop dynamics for such systems. 
% These closed-loop dynamics are again linear with Gaussian process noise and can be determined prior to planning. 
The authors then apply convex optimization techniques to solve the closed-loop, chance-constrained planning problem similar to the algorithm proposed in \cite{Blackmore2009}. 
They also briefly outline a method on how to treat collision avoidance in a multi-agent setting by manual partition of each agent's feasible region prior to planning in such a way that collisions are avoided. 
This need for manual tuning can be a disadvantage since it has to be done for each new instance of a multi-agent problem and it cannot be guaranteed that the partitions are optimal. 

The dissertation \cite{DuToit2009} also accounts for possible future measurements during computation of the control strategy and extends their algorithm to model uncertain and possibly dynamic environments.  % with chance constraints on obstacle collisions. 
Although the results of this work allow to model interactions among several agents with uncertain localization, the computation of a control strategy is limited to a single agent while the other agents are treated as moving obstacles. 
% TODO: Hier mehr????
In~\cite{Cinquemani2009} the authors give a comparison of the \emph{conservative constraints} from~\cite{Hessem2002} and the constraints for \emph{Gaussian} disturbance employing constraint separation from~\cite{Blackmore2009}. 

In \cite{Blackmore2010}, the authors propose to approximate the probabilities for collisions with obstacles for the single agent case by \emph{sample-based Monte-Carlo methods}. 
This approximation has the advantage that almost arbitrary state and noise distributions can be treated. 
They transform the search for an optimal solution to the stochastic single agent control problem under chance constraints to finding a solution of a mixed integer linear program (MILP). 
The formulation as a MILP has the advantage that standard solvers like CPLEX \cite{Cplex2008} can be employed to solve problems of moderate size efficiently to an optimal solution. 
However, the considerations in \cite{Blackmore2010} are restricted to chance-constrained control of a single agent.

% {\bf Structure of the paper}
% {\bf Related Work}
\subsection{Notation and Conventions}
% \begin{itemize}
Bold face letters $\rvec{X}$ or $\rvec{x}$ denote random variables, where the underline indicates that the random variable is multivariate. 
Bold face letters $\mat{A}$, $\mat{B}$, or $\mat{C}$ denote matrices. 
A superscript ${}^T$ denotes the transpose of a vector or a matrix. 
Deterministic quantities such as the system input $\vec{u}^i_t$ will be in normal type, where the underline indicates that the variable is a vector. 
The first superscript of a variable identifies the agent this variable refers to. 
The first subscript identifies the time instance in the planning horizon. 
The expectation of a function $g$ with respect to a random vector $\rvec{X}$ with probability density $f(\vec{x})$ is defined as $\erw[\rvec{X}]{g} := \int g(\vec{x}) f(\vec{x}) d \vec{x}$. 
Similarly, the expectation of a general event $E$ with respect to a random vector $\rvec{X}$ with probability density $f(\vec{x})$ is defined as $\erw[\rvec{X}]{E}:= \int_E f(\vec{x}) d \vec{x} = \int \chi_E(\vec{x}) f(\vec{x}) d \vec{x}$, where $\chi_E(\vec{x})$ is the indicator function of $E$. 
The probability that a multivariate random variable $\rvec{x}$ with probability density function (pdf) $f(\vec{x})$ lies in some set $E$ is denoted by $\Pr(\rvec{x} \in E)$ and is defined as the multivariate expectation
\begin{align}
 \Pr(\rvec{x} \in E) := \int_E f(\vec{x}) d \vec{x} = \int \chi_E(\vec{x}) f(\vec{x}) d \vec{x} \ ,
\end{align}
where $\chi_E$ is the indicator function of $E$, ie. $\chi_E(\vec{x})$ equals one if $\vec{x} \in E$ and zero otherwise. 
$\Cov(\rvec{X})$ denotes the covariance $\erw{(\rvec{X} - \erw{\rvec{X}}) (\rvec{X} - \erw{\rvec{X}})^T}$ of the random vector $\rvec{X}$, which is a matrix for multivariate random variables. 
Analogously $\Cov(\rvec{X}, \rvec{Y}) := \erw{(\rvec{X} - \erw{\rvec{X}}) (\rvec{Y} - \erw{\rvec{Y}})^T} $ is the covariance between the random vectors $\rvec{X}$ and $\rvec{Y}$ \cite{Papoulis2002}. 
The Manhattan-norm is defined through $\Vert \vec{x} \Vert_1 := \sum_{k} \vert x_k \vert$, the Euclidean norm through $\Vert \vec{x} \Vert_2:= \sqrt{\sum_k x_k^2}$ and the sup-norm as $\Vert \vec{x} \Vert_{\infty} := \max_{k} \vert x_k \vert$ for real vectors $\vec{x} = [x_1, x_2, \dots, x_n]^T$.

\section{General Problem Formulation}\label{sec:gen_prob}

% \begin{itemize}
% \subsection{Problem Formulation}
The general problem we want to solve is as follows:
For $M$ agents $i = 1,\dots,M$ with discrete-time stochastic description in state space form, we plan over a horizon of length $H$ in order to minimize the sum of the agents' cost functions. 
This minimization is subject to the constraints that the probabilities of inter-agent collisions and the probability of agents leaving the feasible region are kept below certain user-defined thresholds. 
The formal formulation of this problem is
% \end{itemize}
% \begin{equation}
  \begin{align}
      \underset{\vec{u}^i_{1:H}, i = 1, \dots, M}{\mathrm{minimize}} & \   \sum_{i = 1}^M \erw[\rvec{x}^i_{0:H}]{h^i(\vec{x}^i_{0:H}, \vec{u}^i_{1:H})} \label{eq:objective} \\
			    \mathrm{s.t. }  \ & \forall_{i = 1, \dots, M} \  \vec{u}^i_{1:H} \in F^i_u \\ &
  			    \forall_{i = 1, \dots, M, t = 1, \dots, H} \  \rvec{x}^i_t = f^i_t(\rvec{x}^i_{0}, \vec{u}^i_{1:t}, \rvec{\nu}^i_{1:t}) \\ &  
			    \forall_{i = 1, \dots, M} \  \Pr(\rvec{x}^i_{1:H} \notin F^i) \le \delta^i \label{eq:c_constraint}\\ &
			    \forall_{\underset{j \neq i}{i,j = 1, \dots, M,}} \ \Pr((\rvec{x}^i_{1:H}, \rvec{x}^j_{1:H}) \notin F^{i,j}) \le \delta^{i,j} \label{eq:ca_constraint} \ .
  \end{align}
% \end{equation}
% \begin{itemize}
% This is the stochastic model predictive control problem with chance constraints on the failure of the mission. 
The decision variables $\vec{u}^i_{1:H} = [(\vec{u}^i_1)^T, (\vec{u}^i_2)^T, \dots , (\vec{u}^i_H)^T]^T$ are the (deterministic) control inputs to agent $i$ confined to lie in the polygonal and convex feasible region  $F^i_u$. 
The function $h^i$ is the control objective and rates how desirable certain states of the agents are. 
It depends on the control inputs and the system states of the agents. 
The system state of agent $i$ over the planning horizon is modeled as a random vector and is denoted by $\rvec{x}^i_{0:H} = [(\rvec{x}^i_0)^T, (\rvec{x}^i_1)^T, \dots, (\rvec{x}^i_H)^T]^T$. 
We assume these random vectors to be stochastically independent for different agents $i_0 \neq i_1$ and we also assume that the system noise is independent of the control inputs. 
Since the states are modeled as random vectors, we take the expectation of $h^i(\vec{x}^i_{0:H}, \vec{u}^i_{1:H})$ w.r.t. the agents' state distributions in (\ref{eq:objective}). 

The mapping $f^i_t$ describes the model of the dynamics of agent $i$. 
The state of agent $i$ at time instance $t$ depends on the initial state $\rvec{x}^i_0$, the control inputs $\vec{u}^i_1, \dots, \vec{u}^i_{t}$, and stochastic noise $\rvec{\nu}^i_1, \dots, \rvec{\nu}^i_t$ acting upon each agent. 
The stochastic noise terms are used to account for possible errors in the dynamic model or exogenous disturbances that act upon the systems, such as wind turbulence on UAVs.
They can also account for uncertainties in the initial state estimate $\rvec{x}^i_0$ and how those are carried forward and possibly increased through state prediction over time. 
We assume here that the second-order central moments of the noise terms $\rvec{\nu}^i_t$ and the prior state distributions $\rvec{x}^i_0$ are known. 
Even if there is no analytic knowledge about the second-order moments, since we will assume in later sections that we can draw samples from these distributions, so it should be possible to obtain good estimates of these quantities through the sample approximations. 

$F^i$ is the feasible region to agent $i$ and $\Pr(\rvec{x}^i_{1:H} \notin F^i)$ is the probability that agent $i$ leaves the feasible region during the mission. 
The feasible region can for example model an area the agents are not supposed to leave or obstacles the agents have to avoid. 
$\Pr((\rvec{x}^i_{1:H} , \rvec{x}^j_{1:H}) \notin F^{i,j})$ specifies the probability that agent $i$ and agent $j$ do not meet the coupling constraints defined by the feasible region $F^{i,j}$ that control the interaction among agents. 
We consider constraints consistent for all agents, so we set $F^c := F^{i,j}$. 
We understand these constraints as collision avoidance constraints, i.e., constraints on the expected distance of agent $i$ and $j$. 

The upper bounds $0 \le \delta^i, \delta^{i,j} \le 1$ on the probability of the agents leaving their own feasible region $F^i$ or the joint feasible regions $F^c$ characterize the chance constraints on the failure of planning \cite{Schwarm1999,Birge1997,Prekopa1995}. 
These chance constraint bounds are specified by the user and can be used to adjust how cautious the agents' plans will become. 
\section{Sample-based Methods for Systems with Linear Dynamics}\label{sec:sys_dynamics}
% In this section we will give the model of the system dynamics we will use and
We assume that the stochastic discrete-time dynamic state space model of each agent $i = 1, \dots, M$ is given through the linear system equation
% \end{itemize}
\begin{align}\label{eq:system_dynamics}
 \rvec{x}^i_{t + 1} = \mat{A}^i \rvec{x}^i_t + \mat{B}^i \vec{u}^i_t + \rvec{\nu}^i_t \ ,
\end{align}
where $\rvec{x}^i_t$ is the uncertain state of agent $i$ modeled as a random variable, $\vec{u}^i_t$ is the deterministic control input and $\rvec{\nu}^i_t$ is the stochastic system noise. 
% \begin{itemize}
This model can either be obtained by assuming linear dynamics from the outset or by linearizing the system dynamics around a working point. 
For linear system dynamics of the form~(\ref{eq:system_dynamics}) the mapping 
\begin{align}
\rvec{x}^i_t = f^i_t(\rvec{x}^i_0, \vec{u}^i_{1:t}, \rvec{\nu}^i_{1:t}) \ ,
\end{align}
that determines how the agent's state depends on the initial state $\rvec{x}^i_0$ and behaves under control inputs $\vec{u}^i_{1:t}$, is linear in the control inputs and given through the equation
% \end{itemize}
\begin{align}\label{eq:system_dynamics2}
  \rvec{x}^i_{t + 1} =  (\mat{A}^i)^{t}\rvec{x}^i_{0} + \sum_{s = 1}^{t} (\mat{A}^i)^{t -s -1}(\mat{B}\vec{u}^i_{s} + \rvec{\nu}^i_{s}) \ ,
\end{align}
as can be easily checked. 
% % \begin{itemize}
For arbitrarily distributed initial states $\rvec{x}^i_0$ and for arbitrarily (non-Gaussian) distributed system noise $\rvec{\nu}^i_t$ there is in general no closed-form representation (or one with a finite number of parameters) of the random vectors $\rvec{x}^i_t$. 
We therefore make use of sample-based methods to represent the uncertain states of the agents and will describe how to do so in the rest of this section. 

We assume that for each agent $i = 1, \dots, M$ we can draw $N$ independent samples distributed with the same distribution as the agent's state $\rvec{x}^i_0$ at time step $t = 0$. 
We will denote these samples by $\{\vec{x}^i_{0,j} \}_{j=1}^N$ and assume, % where $\vec{x}^i_{0,j}$ are the positions of the samples and $\omega^i_{0,j}$ the importance weights, normalized to sum up to one. 
for notational convenience only, that we draw an equal number $N$ of samples from each agent's prior distribution. 
Furthermore for each agent $i$ we draw $N$ noise samples 
\begin{align}
(\vec{\nu}^i_{1,j},\vec{\nu}^i_{2,j}, \dots, \vec{\nu}^i_{H,j}) \sim f^{\vec{\nu}^i_{1:H}}(\vec{\nu}^i_{1},\vec{\nu}^i_{2}, \dots, \vec{\nu}^i_{H}) \ , \ j = 1, \dots, N
\end{align}
from the distribution of the system noise $\rvec{\nu}^i_1, \dots, \rvec{\nu}^i_H$ that affects agent $i$ over the planning horizon of length $H$.

The model of the dynamics (\ref{eq:system_dynamics}) of each agent $i$ allows us now to generate $N$ 'sample trajectories' over time.  
These trajectories are obtained by propagating each initial sample in combination with a noise sample through the system equation~(\ref{eq:system_dynamics2}). 
Each sample trajectory has the length $H+1$ and consists of samples 
\begin{align}
  \vec{x}^i_{0:H,j} := [(\vec{x}^i_{0,j})^T, (\vec{x}^i_{1,j})^T, \dots, (\vec{x}^i_{H,j})^T]^T \ . 
\end{align}
The ($t+1$)-th sample is defined through the formula
\begin{align}\label{eq:rec_law}
 \vec{x}^i_{t + 1,j} =  (\mat{A}^i)^{t}\vec{x}^i_{0,j} + \sum_{s = 1}^{t} (\mat{A}^i)^{t -s -1}(\mat{B}\vec{u}^i_{s} + \vec{\nu}^i_{s,j})  \ .
\end{align}
% and will be denoted by $\vec{x}^i_{0:H,j} := [(\vec{x}^i_{0,j})^T, (\vec{x}^i_{1,j})^T, \dots, (\vec{x}^i_{H,j})^T]^T$.
% \end{itemize}
% \begin{align}\label{eq:rec_sample}
%   \vec{x}^i_{t,j} = f^i_t(\vec{x}^i_{0:t-1}, \vec{u}^i_{1:t}, \vec{\nu}^i_{t,j} ) 
% \end{align}
% \begin{itemize}
The most important property of this construction is that the 'sample trajectories' depend deterministically and linearly upon the control inputs. %, after the disturbance realizations $\vec{\nu}^i_{t,j}$ are drawn. 
This will enable us to formulate the sample-based approximation stochastic control problem as a deterministic optimization problem. 

% For every time instance $t$ in the planning horizon and for each agent $i$, (\ref{eq:rec_law}) allows to generate a sample representation $\{ \vec{x}^i_{t,j} \}_{j = 1}^N$ that depends deterministically on all the previous control inputs $\vec{u}_{1:t}$ up to $t$. 
% Through prediction of the samples over time with linear system dynamics, the weights $\omega^i_{0,j}$ do not change. 
% This sample representation $\{ \vec{x}^i_{t,j} \}_{j = 1}^N$ is an approximation of the underlying probability density function $f^i_t$ of the random vector $\rvec{x}^i_t \sim f^i_t$. 
% For linear systems the recursive law (\ref{eq:rec_sample}) corresponds to 
% % \end{itemize}
% \begin{align}
%  \vec{x}^i_{t + 1,j} =  (\mat{A}^i)^{t}\vec{x}^i_{0,j} + \sum_{s = 1}^{t} (\mat{A}^i)^{t -s -1}(\mat{B}\vec{u}^i_{s} + \vec{\nu}^i_{s,j}) \ .
% \end{align}
% \begin{itemize}
So what have we achieved by this construction? 
For one we have obtained a representation of agent $i$'s uncertain states for instance at time step $t_0$ the samples $\{ \vec{x}^i_{t_0,j} \}_{j=1}^N$ approximate the distribution of the random variable $\rvec{x}^i_{t_0}$. 
This approximation can be used to evaluate probabilistic quantities depending on the agents' state distributions that otherwise we would have had to employ computationally expensive methods to calculate. 
One example of such a probabilistic quantity is the mean of the uncertain state of agent $i$ at some time step $t_0$ which we can now estimate by the sample mean: 

The sample mean vector 
\begin{align}
  \widehat{\vec{\mu}}^i_{t_0}:= \frac{1}{N}  \sum_{j=1}^N \vec{x}^i_{t_0,j}
\end{align}
 %and the sample covariance matrix 
% \begin{align}
% \widehat{\mat{C}}^i_t := \frac{\sum_{j = 1}^N \omega^i_{0,j}(\vec{x}^i_{t,j} - \vec{\mu}^i_t)(\vec{x}^i_{t,j} - \vec{\mu}^i_t)^T }{1 - \sum_{j = 1}^N \omega^i_{0,j}}
% \end{align}
is an estimator of the true first moment of $\rvec{x}^i_{t_0}$, with convergence to the true value as the number of samples goes to infinity.
As another example the expectation $\erw[\rvec{x}^i_{t_0}]{g}$ can be approximated through
\begin{align}
  \erw[\rvec{x}^i_{t_0}]{g} \approx \frac{1}{N} \sum_{j=1}^N  g(\vec{x}^i_{t_0,j})
\end{align}
and the expectation over an event $E$ through
\begin{align}
  \erw[\rvec{x}^i_{t_0}]{E} \approx \frac{1}{N} \sum_{j=1}^N \chi_E(\vec{x}^i_{t_0,j})
\end{align}
both with convergence to the true expectation when the number of samples goes to infinity. % $N \to \infty$.

We will use these sample-based approximations throughout the rest of the paper to approximate probabilities of agents leaving the feasible region or colliding with each other. 

\section{Sample-based Approximation of Obstacle Collision Probabilities}\label{sec:sa_milp}

% {\bf Bild von obstacle und Partikel liegen dr\"uber, links, oder rechts davon w\"are in dieser Section gut}
In this section, we will outline how a tractable approximation of the single agent chance-constrained MPC problem can be formulated, using the sample approximation of an agent's state distribution from the previous section. 
The next section will be concerned with modeling and approximating the coupling constraints between the agents. 
Since we will only consider the single agent case in this sectio, we will drop the superscript denoting the agent for notational convenience for the rest of the section. 

In the work \cite{Blackmore2010}, the authors showed that for a single agent with linear dynamics and a sample-based approximation of the agent's state and under further assumptions, the chance-constrained MPC problem can be formulated as a mixed integer linear programm.
The additional assumptions are that the objective function $h$ is piecewise linear and convex and the feasible region $F$ is either convex and polygonal or non-convex with polygonal obstacles. %, the single agent chance-constrained MPC problem can be formulated as a MILP.
% We will now describe how to formulate the program for a piecewise linear convex objective function $h^i$ and a non-convex feasible region $F^i$ that corresponds to $\mathbb{R}^2$ with a single axis-aligned rectangular obstacle $O$.
In the following we will outline the construction from \cite{Blackmore2010}. 

If $h$ is a piecewise linear convex objective function, for example a function consisting of weighted Manhattan-norms (for an example see for instance~\cite{Schouwenaars2001} or the simulations in Section~\ref{sec:sim}), the sample approximation of the expectation of the cost function
\begin{align}\label{eq:plc}
  \erw[\rvec{x}_{0:H}]{h(\vec{x}_{0:H}, \vec{u}_{1:H})} \approx \frac{1}{N} \sum_{j=1}^N  h(\vec{x}_{0:H,j} \ , \vec{u}_{1:H})
\end{align}
%  the expectation of the cost function taken over $\rvec{x}_{0:H}$ approximated by the sample expectation
is again a piecewise linear convex function in the control parameters $\vec{u}_{1:H}$. 
This can be seen as follows. 
By construction we know that the sample trajectories depend linearly upon the control inputs. 
The composition of this linear mapping with the piecewise linear and convex function $h$ is again piecewise linear and convex. 
Since taking the expectation of this composition in Eq.~(\ref{eq:plc}) is a linear operation, the whole sample expectation of the objective function is piecewise linear and convex.

The probability that the agent is not in the feasible region $F$ at time instance $t$ is approximated through the sample expectation 
\begin{align}
\Pr(\rvec{x}_{1:H} \notin F) = \erw[\rvec{x}_{1:H}]{\chi_{CF}} \approx \frac{1}{N} \sum_{j=1}^N  \chi_{CF}(\vec{x}_{1:H,j}) \ , 
\end{align}
where $\chi_{CF}$ is the indicator function of the complement $\mathbb{R}^n \setminus F$ of the feasible region. 
It is defined to be $\chi_{CF}(\vec{x}) = 1$ if $\vec{x}$ is not in the feasible region (and hence in the complement $\mathbb{R}^n \setminus F$) and $\chi_{CF}(\vec{x}) = 0$ if $\vec{x}$ is in the feasible region $F$. 
Instead of the true, computationally intractable, chance constraint $\Pr(\rvec{x}_{1:H} \notin F) \le \delta$ we impose the chance constraint on the approximated probability
\begin{align}\label{eq:sa_approx_cc}
  \frac{1}{N} \sum_{j=1}^N  \chi_{CF}(\vec{x}_{1:H,j})  \le \delta \ .
\end{align}
This constraint on the approximated probability can be evaluated much more efficiently and can also be transformed into mixed integer and linear constraints on the optimization problem. 
This can be achieved by setting constraints on the number of samples that are not within the feasible region $F$ as we will now outline. 

For this purpose, the function that indicates whether a particle trajectory leaves the feasible region $\chi_{CF}(\vec{x}_{1:H,j})$ is replaced by a binary variable $e_{j} \in \{0,1\}$.
This binary variable is equal to one $e_{j} = 1$ if the $j$-th sample trajectory leaves the feasible region and equal to zero $e_{j} = 0$ if the sample trajectory stays within the feasible region. 
So how is this binary variable defined? 
For a convex polygonal feasible region, $e_{j}$ evaluates whether sample trajectory $\vec{x}_{1:H,j}$ fulfills all linear inequalities given by the line segments defining $F$. 
For polygonal obstacles in the feasible region, $e_{j}$ measures whether certain integer constraints are satisfied that specify whether the sample trajectory $\vec{x}_{1:H,j}$ has positive clearance to every face of the polygonal obstacle. 
So in general $e_{j}$ measures if the $j$-th particle trajectory satisfies all constraints of a set of linear constraints and a set of mixed integer linear constraints. 
The constraint~(\ref{eq:sa_approx_cc}) that only a weighted subset of all particles leave the feasible region is then transformed into the integer constraint
\begin{align}
  \frac{1}{N} \sum_{j=1}^N  e_{j} \le \delta \ 
\end{align}
that only a weighted subset of the binary variables $e_j$ is equal to one. 

The MILP formulation of the single agent chance-constrained MPC problem is then given by
% \small
\begin{equation}
  \begin{aligned}
   \underset{\vec{u}_{1:H}}{\mathrm{minimize}} \  & \frac{1}{N} \sum_{j=1}^N  h(\vec{x}_{0:H,j}) \\
    \mathrm{s. t.}\ 	 & \vec{u}_{1:H} \in F_u \\
			 & \vec{x}_{t,j} = f^i_t(\vec{x}_{0,j}, \vec{u}_{1:t}, \vec{\nu}_{1:t,j}) \\ 
			 & \frac{1}{N} \sum_{j=1}^N  e_{j} \le \delta \\ 
			 & e_{j} \in \{0,1\} \text{ determined from feasible region. } 
  \end{aligned}
\end{equation}
For more detailed descriptions on how to derive the mixed integer linear constraints from the feasible region please refer to \cite{Schouwenaars2001} and \cite{Blackmore2010}.

\section{Multi-Agent Collision Avoidance}\label{sec:col_agents}

% A probabilistic formulation of collision avoidance with other agents can be constructed as MILP constraints similar to obstacle avoidance.
% The drawback of this formulation is that the number of binary variables explodes, as will be seen in later sections, and, hence, the solution time of the program becomes to long and computationally demanding to be feasible for real world scenarios.
% We therefore propose a computationally much more attractive formulation and show how this formulation still enables us to bound collision probabilities. 
% The formulation relies relies on the fact that in the case of linear system dynamics like the ones in eq.~\ref{eq:model}, the covariance of the state distribution does not depend on the actual control inputs applied.
This section is concerned with the approximation of the probabilities of inter-agent collisions and the derivation of constraints that keep these probabilities below user-defined thresholds. 
We will first formally define our notion of an inter-agent collision (the agents come too close) and then study how from this definition, the true probability of an inter-agent collision can be derived. 
Because the probability of a collision of two agents will depend on \emph{both} uncertain states of the agents its evaluation is even more computationally expensive than the probability of a single agent leaving the feasible region. 
So in a first step we use the sample approximations of the agents' uncertain states to derive a sample approximation of the probability of an inter-agent collision. 
This direct approximation will converge to the true probability of a collision as the number of samples goes to infinity, however, the complexity of the resulting optimization problem can be so high that this approach proves problematic for control under real-time requirements. 
Because of its convergence, however, solutions found with this formulation will be very close to the optimal solutions of the ``true'' chance-constrained MPC without any approximation. 
So the complexity of this formulation calls for other more tractable ones but it will serve as a benchmark for the others.

For simplicity of exposition and for notational convenience we will consider path planning in the two-dimensional plane and we will from now on only consider two agents, denoted by a superscript $1$ and $2$. 
Since we defined the constraints on the probability of a collision in~(\ref{eq:ca_constraint}) as pairwise constraints on the states of two agents, the considerations can be extended verbatim for every other possible combination of two agents. 
% We will enforce the chance constraint on the probability of a collision of the two agents at some time instance $t$ in the planning horizon. %the chance constraint for all time instances can be formulated as a constraint on the sum of these constraints. 

\subsection{Sample-based Approximation of Inter-agent Collision Probability}\label{sec:prob_col_av}
In this section we will define the event of a collision of two agents, study the probability of such an event occurring depending on the uncertain states of the agents and will derive a sample-based approximation of this probability. 

Let $\epsilon > 0$ be the prespecified minimum distance between two agents for collision avoidance, for instance we could set $\epsilon$ to be twice the diameter of a robot or the twice the wingspan of a fixed-wing UAV. 

We let $\vec{x}^{1}_t$ and $\vec{x}^{2}_t$ denote the two-dimensional positions of agents $1$ and $2$ in the plane at a time instance $t$. % $1$ and $2$ at some time instance $t$. % and let $\Vert \cdot \Vert_2$ denote the two-norm in Euclidean space. 
We define the event of a \emph{collision} as $\Vert \vec{x}^{1}_t - \vec{x}^{2}_t \Vert_2 < \epsilon$, i.e., the two agents' are closer than the minimum clearance $\epsilon$. 
 % $ \Vert \vec{x}^{1}_t - \vec{x}^{2}_t \Vert_1 < \epsilon$ we also have  $\Vert \vec{x}^{1}_t - \vec{x}^{2}_t \Vert_2 < \epsilon$.
The feasible region $F^c$ for the joint states of the agents resulting from this definition is the set of possible positions that do not cause a collision. 
Hence the feasible set is the set of agent positions that have a distance greater or equal than $ \epsilon$
\begin{align}
  F^c:= \{ (\vec{x}^{1}, \vec{x}^{2}) \vert \Vert \vec{x}^{1} - \vec{x}^{2} \Vert_{2} \ge  \epsilon \} \ . 
\end{align}
% If the uncertain descriptions of the agents states are roughly understood as weighted position, with higher weight when a higher probability of presence is assumed, the probability of a collision 
The probability of a collision of two agents is then the probability that the uncertain states of the agents are not in the feasible set
\begin{align}
 \Pr &((\rvec{x}^{1}_t , \rvec{x}^{2}_t) \notin F^c)  = \Pr(\Vert \rvec{x}^{1}_t - \rvec{x}^{2}_t \Vert_{2} < \epsilon) \\  
  & = \erw[\rvec{x}^{1}_t ,\ \rvec{x}^{2}_t]{\chi_{CF^c}(\vec{x}^{1}_t,\vec{x}^{2}_t)} \\
  & = \int \int \chi_{CF^c}(\vec{x}^{1}_t, \vec{x}^{2}_t) f(\vec{x}^{1}_t)f(\vec{x}^{2}_t)d\vec{x}^{1}_t d\vec{x}^{2}_t \label{eq:ma_ca} \ ,
\end{align}
where
\begin{align}\label{eq:def_chi_c}
  \chi_{CF^c}(\vec{x}^{1}_t,\vec{x}^{2}_t) = \begin{cases}
                        1 & \mathrm{ if } \  \Vert \vec{x}^{1}_t - \vec{x}^{2}_t \Vert_2 < \epsilon  \\
                        0 & \mathrm{ else }
                       \end{cases} 
\end{align}
is the indicator function of the complement of $F^c$ and $f(\vec{x}^{1}_t)$ and $f(\vec{x}^{2}_t)$ are the probability density functions of the position estimates of the agents. %which are assumed to be stochastically independent. 
Please note that in the above we have used the relation $(\vec{x}^1, \vec{x}^2) \notin F^c \Leftrightarrow \Vert \vec{x}^1 - \vec{x}^2 \Vert_2 < \epsilon$ which follows from the definition of the feasible set $F^c$. 

The integral~(\ref{eq:ma_ca}) that determines the probability of a collision is the integral of multivariate density functions multiplied by the indicator function $\chi_{CF^c}$. 
For general density functions this integral will have no closed-form solution because of the possibly complex structure of the density functions. % and non-linearity of the indicator function. 
Even for multivariate Gaussian distributions it will become difficult to evaluate the integral since the indicator function is a non-convex and nonlinear function and there are no known formulas to evaluate this integral.  
%
% It is comprised of an integration over two multivariate distributions $f(\vec{x}^{1}_t)$ and $f(\vec{x}^{2}_t)$. 
% In general it can not be assumed, that the multivariate distributions over the agents' respective state estimates $f(\vec{x}^{1}_t)$ and $f(\vec{x}^{2}_t)$ are independent in the dimension of the state estimate, i.e.,. $f(\vec{x}^{1}_t) = f(x^{1}_t) \cdot f(y^{1}_t)$, where $\vec{x}^{1}_t = [x^{1}_t, y^{1}_t]^T$ is the two-dimensional position estimate, and analogously for $2$.
% Thus the integration can only be simplified into one-dimensional integrals through the conservative bounding assumption of independence of the state estimates. 
Even if the integral (\ref{eq:ma_ca}) was given by a closed-form representation, it is not guaranteed that the resulting constraints on the probability of an inter-agent collision would be tractable for an optimization algorithm. %to be a convex function in the joint control inputs of the two agents and, hence, it would not be clear if the chance constraints could be formulated as convex constraints. 

In order to make the probability of an inter-agent collision computationally more tractable, we approximate the probability a collision through the sample expectation. 
% Sample approximations of the agents' state distributions are a way to approximate the probability of an inter-agent collision and derive a first MILP formulation of the inter-agent collision avoidance chance constraint. 
We do this by replacing the continuous distributions $f(\vec{x}^1_t)$ and $f(\vec{x}^2_t)$ through their corresponding sample approximations, given through the two sets of samples $\{\vec{x}^1_{t,j} \}_{j=1}^N$ and $\{ \vec{x}^2_{t,l} \}_{l = 1}^N$. 
When the continuous distributions are replaced by their sample 'counterparts' the double integrals in Equation~(\ref{eq:ma_ca}) are replaced by a nested sum and, hence, the approximated probability of a collision is
\begin{equation}\label{eq:pr_crash}
  \begin{aligned} 
  \Pr & ((\rvec{x}^{1}_t , \rvec{x}^{2}_t) \notin F^c) = \erw[\rvec{x}^{1}_t , \rvec{x}^{2}_t]{\chi_{CF^c}(\vec{x}^{1}_t , \vec{x}^{2}_t)} \\
      & \approx \frac{1}{N^2} \sum_{j = 1}^N \sum_{l = 1}^N   \chi_{CF^c}(\vec{x}^{1}_{t,j},\vec{x}^{2}_{t,l}) \ .
  \end{aligned}
\end{equation}
% where we have replaced the continuous distributions $f(\vec{x}^1_t)$ and 
Results on Monte-Carlo sampling methods show that this approximation will convergence against the true collision probability as the number of samples for each agent goes to infinity (see e.g. \cite{Caflisch1998} for a discussion of convergence properties). 

In the next section we will show how this approximation can be translated into mixed integer linear constraints. 
Unfortunately the complexity of the resulting optimization problem will be so high, that it is difficult to solve the problem with these constraints under real-time requirements. 
% This is the reason why we will propose an alternative method for enforcing constraints on the probability of inter-agent collisions in Section~\ref{sec:col_av_app}. 
However, since the sample-based approximation introduced in this section will converge against the true probability of a collision of two agents it is still highly relevant and it will serve as a benchmark for the alternative approach we will propose. 

% This sample approximation can be used to formulate MILP constraints as we will show in the next section. 
% In the next subsection we will formulate the approximation (\ref{eq:pr_crash}) as MILP constraints. 
% However, the drawback of this direct approximation of the collision avoidance chance constraints is that it leads to an explosion of binary variables as we will show in Section~\ref{sec:bad_comp_ana}. 
%, since we would have to consider each pairing of samples of two agents at every time step. 

\subsection{Constraint Formulation for Sample-based Approximations}\label{sec:bad_constraints}
In this section, we will outline how the sample approximation~(\ref{eq:pr_crash}) of the probability of a collision between two agents can be transformed into mixed integer linear constraints. 
% This section is rather technical and can be skipped for a first reading. 

As a first step we have to replace the Euclidean norm in the definition of the feasible set 
\begin{align}
F^c = \left\{ (\vec{x}^{1}_t, \vec{x}^{2}_t) \vert \ \Vert \vec{x}^{1}_t - \vec{x}^{2}_t \Vert_{2} \ge \epsilon \right\}
\end{align}
by the supremum norm, since the Euclidean norm $\Vert \cdot \Vert_2$ involves evaluation of quadratic terms that cannot be transformed into mixed integer linear constraints. %, we will replace it with the supremum norm. 
So we replace the original feasible $F^c$ set by the auxiliary set
\begin{align}
  F^c_{\infty} := \left\{ (\vec{x}^{1}_t, \vec{x}^{2}_t) \vert \ \Vert \vec{x}^{1}_t - \vec{x}^{2}_t \Vert_{\infty} \ge \epsilon \right\} \ .
\end{align}

Replacing the Euclidean norm by the supremum only makes the feasible set for two agents smaller because from the norm inequality $\Vert \vec{x}^{1}_t - \vec{x}^{2}_t \Vert_{\infty} \le \Vert \vec{x}^{1}_t - \vec{x}^{2}_t \Vert_{2}$, it follows that if two agents have a distance of greater or equal than $\epsilon$ in the supremum norm, they have a distance greater or equal than $\epsilon$ in the Euclidean norm. 
So it follows that $F^c_{\infty}$ is a subset of $F^c$ and hence, we only make the feasible region for the agents smaller and are more conservative when considering the supremum norm instead of the Euclidean norm. 
In this section only, we will consider the feasible region $F^c_{\infty}$. %will be understood as the set of points that have distance greater or equal $1 \slash \sqrt{2}\epsilon$ in the supremum-norm. %instead of $\epsilon$ in the Euclidean norm. 

According to (\ref{eq:pr_crash}), the approximation of the constraints $\Pr((\rvec{x}^{1}_t , \rvec{x}^{2}_t) \notin F^c_{\infty}) \le \delta^{1,2}_t$ is %in the sample approximation as shown above in (\ref{eq:pr_crash})
\begin{align}
 \frac{1}{N^2} \sum_{j = 1}^N \sum_{l = 1}^N  \chi_{CF^c_{\infty}}(\vec{x}^{1}_{t,j},\vec{x}^{2}_{t,l}) \le \delta^{1,2}_t \ ,
\end{align}
where
\begin{align}
 \chi_{CF^c_{\infty}} (\vec{x}^{1}_{t,j},\vec{x}^{2}_{t,l}) := \begin{cases}
                                                                       1,\ & \text{ if } \Vert \vec{x}^{1}_{t,j} - \vec{x}^{2}_{t,l} \Vert_{\infty} <  \epsilon \\
								       0, \ & \text{ otherwise,}
                                                                      \end{cases}
\end{align}
is the indicator function of the complement of $F^c_{\infty}$. 

To evaluate the sum above, we have to iterate through all samples $j= 1\dots N$ of the first agent and all samples $l = 1 \dots N$ of the second agent and check if $\chi_{CF^c}(\vec{x}^{1}_{t,j}, \vec{x}^{2}_{t,l})$ equals one or zero. 
In order to formulate this 'check' we introduce a binary variable $e_{j, l, t, 1, 2} \in \{ 0, 1 \}$ for that holds if $e_{j, l, t, 1, 2} = 1$ then samples $\vec{x}^{1}_{t,j}$ and $\vec{x}^{2}_{t,l}$ are within $\epsilon$ proximity of each other and if $e_{j, l, t, 1, 2} = 0$ they are not. 
Then we have to ensure that the weighted sum of the $e_{j, l, t, 1, 2}$ is less or equal than the chance constraint bound $\delta^{1,2}_t$. 

So how do we construct the binary variable $e_{j, l, t, 1, 2}$? 
By definition, the sample $\vec{x}_{t,j}^{1} = [x_{t,j}^{1}, y_{t,j}^{1}]^T$ is in more than $\epsilon$ away from sample $\vec{x}_{t,l}^{2} = [x_{t,l}^{2}, y_{t,l}^{2}]^T$ in the supremum norm if
\begin{align}\label{eq:norm_eps}
\Vert \vec{x}_{t,j}^{1} - \vec{x}_{t,l}^{2} \Vert_{\infty} = \max\{ \vert x_{t,j}^{1} - x_{t,l}^{2} \vert\ , \vert y_{t,j}^{1} - y_{t,l}^{2} \vert \} > \epsilon \ .
\end{align}
This is equivalent to the condition that one of the following inequalities holds
\begin{align}
 &  x_{t,j}^{1} - x_{t,l}^{2} >  \epsilon \text{ or } \\ &
    x_{t,j}^{2} - x_{t,l}^{1} >  \epsilon \text{ or } \\ &
    y_{t,j}^{1} - y_{t,l}^{2} >  \epsilon \text{ or } \\ &
    y_{t,j}^{2} - y_{t,l}^{1} >  \epsilon \text{ or } \ ,
\end{align}
where we just formulated the absolute value from Eq.~\ref{eq:norm_eps} differently. 
Since the logical 'or'-constraints above are not directly applicable as integer linear constraints, we use the 'Big M'-method to transform them into logical 'and'-constraints. 

So we check whether sample $\vec{x}_{t,j}^{1} = [x_{t,j}^{1}, y_{t,j}^{1}]^T$ is in $\epsilon$ proximity of sample $\vec{x}_{t,l}^{2} = [x_{t,l}^{2}, y_{t,l}^{2}]^T$ at time instance $t$ through the constraints
% \small
\begin{align}
 & x_{t,j}^{1} - x_{t,l}^{2} >  \epsilon - b^1_{j,l,t,1,2}M_o \text{ and } \label{eq:milp1}\\&
   x_{t,l}^{2} - x_{t,j}^{1} >   \epsilon - b^2_{j,l,t,1,2}M_o \text{ and } \\&
   y_{t,j}^{1} - y_{t,l}^{2} >   \epsilon - b^3_{j,lt,1,2}M_o \text{ and } \\&
   y_{t,l}^{2} - y_{t,j}^{1} >  \epsilon - b^4_{j,l,t,1,2}M_o  \label{eq:milp4}  \ ,
\end{align}
% \normalsize 
with additional binary slack variables $b^i_{j,l,t,1,2} \in \{0,1\}$ and arbitrary large positive number $M_o$. 
If at least one of the $b^i_{j,l,t,1,2}$ above is zero, then the samples have sufficient distance because then the corresponding inequality holds. 
If all four $b^i_{j,l,t,1,2}$ equal one, the samples can be within $\epsilon$ distance of each other. 

Now we define the binary variable $e_{j,l, t,1,2}$ in such a way that it counts the number of samples in too close proximity by the constraint
\begin{align}\label{eq:milp5}
 \sum_{i=1}^4 b^i_{j,l,t,1,2} - 3 \le M_o e_{j,l, t,1,2} \ .
\end{align}
We have that if $e_{j,l, t,1,2} = 1$, it follows that all $b^i_{j,l,t,1,2}, \ i=1, \dots, 4$ can be equal to one and the samples $\vec{x}_{t,j}^{1}$ and $\vec{x}_{t,l}^{2}$ can be within $\epsilon$ distance of each other and if $e_{j,l, t,1,2} = 0$, at least one of the $b^i_{j,l,t,1,2}, \ i=1, \dots, 4$ has to be zero and the samples have sufficient distance. 
% If the chance constraint on the probability of a collision of two agents at time instance $t$ is defined to be $\delta^{1,2}_t$, 
Finally, we bound the weighted number of $e_{j,l, t,1,2}$ that are allowed to be equal to one by the constraint
\begin{align}\label{eq:milp6}
 \frac{1}{N^2} \sum_{j = 1}^M \sum_{l = 1}^M  e_{j,l, t,1,2} \le \delta^{1,2}_t \ .
\end{align}
Please note that since the above sum is a nested sum, we have $M^2$ many binary variables $e_{j,l, t,1,2}$. 
This can lead to very slow run times of a MILP with these constraints. 
The constraints (\ref{eq:milp1})-(\ref{eq:milp4}), (\ref{eq:milp5}), and (\ref{eq:milp6}) are the mixed integer linear formulation of the sample approximation of the chance constraint on the probability of a collision between two agents. 

\section{Efficient Collision Avoidance Approximation}\label{sec:col_av_app}
In the previous section we studied the probability of a collision of two agents and saw that it is difficult to compute the true probability a collision even for uncertain states with simple distributions. 
Therefore we proposed a sample-based approximation that will converge against the true probability when the number of samples goes to infinity. 
When we transformed this approximation into mixed integer linear constraints, it turned out that we have to introduce binary variables for each pairing of samples of different agents. 
Hence, the number of binary variables that results from the sample-based approximation can be prohibitively high. 

In this section, we will therefore propose a more efficient formulation of inter-agent collision avoidance constraints based on \emph{regions of increased probability of presence (RIPP)} of agents.
Not only does the RIPP formulation allow us to generate controls under real-time requirements with only a small degree of sub-optimality compared to the optimal solution of the chance constrained problem. 
But we can also formally prove that controls found with the RIPP algorithm are feasible to the original problem with chance constraints on the true probabilities of agent collisions. 
%  and will analytically study the complexity reduction of the optimization problem formed from the RIPP algorithm. 
% than the constraints on the probability in~(\ref{eq:ca_constraint}). %that conservatively bounds $\Pr((\rvec{x}^{1}_t , \rvec{x}^{2}_t) \notin F^c)$ and allows to generate plans for the agents that are feasible for the chance-constrained MPC problem as defined in section~\ref{sec:gen_prob}. 
% We replace the probability of a collision of two agents by a more conservative expression and will formulate constraints to enforce it. 
% Plans for the agents that satisfy the more conservative collision avoidance constraints are feasible for the problem with bounds~(\ref{eq:ca_constraint}) on the probability of a collision. 
% However, the number of binary variables this new formulation introduces to the program is much lower than in the previous formulation. 

The derivation of the RIPP algorithm proceeds in two steps: First we will define a region around the mean value of the uncertain position $\rvec{x}^i$ of an agent $i$ and study the probability that the agent is outside of this region. %certain size around the expected value of the position estimate. 
The intuition is that the larger this region is, the less probable it is that the position of the agent lies outside the region. %, is quantified by a form of the Chebychev inequality. % that holds for boxes as we will define them and for arbitrarily distributed random vectors. 
This intuition can be quantified by a probabilistic inequality that gives an upper bound on the probability that the position of the agent lies outside of the RIPP region. 
% The inequality is a special Chebychev inequality and so the upper bound is based on the covariance of the uncertain position of the agent. 
% holds for arbitrary uncertain states of the agents, regardless of distribution of states. 
% The inequality 
In the second step we introduce constraints to the control problem that ensure that for different agents their respective RIPPs do not overlap. 
We will prove that if these RIPPs have the adequate size and they do not overlap, then we can control the complex collision probabilities in such a way that the do not exceed the predefined bounds. 
In order to define the RIPPs and ensure their non-overlap, we have to do the following two things.
\begin{enumerate}
  \item We have to determine a correct size for the region of increased probability of presence in such a way that just the right amount of probability mass lies outside of the region. We will do this in Section~\ref{sec:RIPP}. 
  \item We have to formulate constraints that ensure that these regions will not overlap for different agents in order to make sure that the probability of a collision does not exceed the chance constraint bound. We will do this in Section~\ref{sec:good_constraints}
\end{enumerate}

As before in Section~\ref{sec:col_agents} we will only consider two agents $1$ and $2$ and their two-dimensional positions in the plane when talking about collision avoidance. 

\subsection{Regions of Increased Probability of Presence (RIPP)}\label{sec:RIPP}
In this section we will construct the region of increased probability of presence for an agent $i$ with uncertain position $\rvec{x}^i$. 
We will omit the subscript $t$ denoting the time step in this section for notational convenience. 
% The considerations in this section are conducted for
% As before in Section~\ref{sec:col_agents} we will only consider two agents $1$ and $2$ and their two-dimensional positions in the plane when talking about collision avoidance. 
 
Let $\vec{\mu}^{i}  := [\mu^{i}_{x},\mu^{i}_{y} ]^T \in \mathbb{R}^2$ be the mean value of the uncertain position $\rvec{x}^{i} $ of agent $i$, where the subscript $x$ and $y$ denote the $x$ and $y$ coordinates in the plane. 
% Again, we will only consider two-dimensional position estimates for collision avoidance in this section mainly for simplicity reasons. 
% The results can be extended to motion planning in higher dimensions when a more general form of the multivariate Chebychev inequality is used. 
%
\emph{We define a rectangular region of increased probability of presence (RIPP) around mean through
%  and $\alpha^{2}  := [ \alpha^{2}_{x},\alpha^{2}_{y}]^T  \in \mathbb{R}^2$
\begin{align}\label{eq:def_box}
  E^{i}  := \{[x^{i} , y^{i} ]^T \big \vert \vert x^{i}  - \mu^{i}_{x} \vert \le \alpha^{i}_{x}, \vert y^{i}  - \mu^{i}_{y} \vert \le \alpha^{i}_{y} \} \subset \mathbb{R}^2\ .
\end{align} 
}
% and
% \begin{align}
%   E^{2}  := \{ \vec{x}^{2}  = [x^{2} , y^{2} ]^T \big \vert \vert x^{2}  - \mu^{2}_{x} \vert \le \alpha^{2}_{x} \text{ and } \vert y^{2}  - \mu^{2}_{y} \vert \le \alpha^{2}_{y} \} \subset \mathbb{R}^2
% \end{align}
The RIPP region $E^i $ describes the set of points in $\mathbb{R}^2$ for which both coordinates deviate at most some distance from the mean value of the uncertain state of the agent. 
Its position depends on the mean of the uncertain state of the agent and its size depends on the two parameters $\alpha^i_x$ and $\alpha^i_y$. 
The reason why the RIPP region is defined to be rectangular since it is very easy and efficient to define mixed integer linear constraints to ensure that two rectangles do not overlap. 
This will become apparent in the next section, when we will formulate constraints that ascertain that these regions do not overlap for differing agents. 

For each agent $i$, we define the probability that the uncertain position of agent $i$ lies outside of the RIPP region $E^i$ as
% that either the $x$ or the $y$ coordinate of the uncertain position $\rvec{x}^i $ deviates more than the parameter $\alpha^{i} $ from the mean value as
\begin{align}\label{eq:cons_prob}
P^{i}  := \Pr(\rvec{x}^i  \notin E^{i} ) = 1 - \Pr(\rvec{x}^i  \in E^i )\ .
\end{align}
% In other words $P^i $ is the probability . 

For larger RIPP regions $E^i $, the probability $P^i $ becomes smaller and smaller as is illustrated by the next theorem from \cite{Whittle1958}. %Chebychev inequality provided in \cite{Whittle1958}. 
\begin{thm}[P. Whittle \cite{Whittle1958}]\label{thm:theorem1}
Let $\rvec{X} = [\rv{x}_1, \dots, \rv{x}_n]^T$ be a zero mean random vector with covariance matrix $\mat{V}$ and define $P:= 1 - \Pr(\vert \rv{x}_j \vert \le \alpha_j;\ j = 1, \dots n)$, then $P \le \mathrm{trace}(\mat{V} \mat{B}^{-1})$, where $\mat{B}$ is any positive definite matrix with diagonal elements $ b_{jj} = \alpha^2_j$. 
For the special case of a bivariate random variable $\rvec{x}^i $ with covariance matrix $(C^i_{kl})_{k,l = 1,2}$ and probability $P^i $ for $\rvec{x}^i $ defined as in~(\ref{eq:cons_prob}), the bound on $P^i $ simplifies to
% \small
\begin{align}\label{eq:chebychev}
  P^i   \le \frac{C_{11} (\alpha^i_{y})^2 +  C_{22} (\alpha^i_{x})^2}{2 (\alpha^i_{x})^2 (\alpha^i_{y})^2}  
	 + \frac{\sqrt{[C_{11} (\alpha^i_{y})^2 + C_{22} (\alpha^i_{x})^2]^2 - 4 C_{12}^2 (\alpha^i_{x})^2 (\alpha^i_{y})^2}}{2 (\alpha^i_{x})^2 (\alpha^i_{y})^2} \ .
\end{align}
% \normalsize
% holds for $P^i $. 
\end{thm}
% 
% It states that
% \begin{align}\label{eq:chebychev}
%   P^i  & \le \frac{c^i_{1} (\alpha^i_{y})^2 +  c^i_{2} (\alpha^i_{x})^2}{2 (\alpha^i_{x})^2 (\alpha^i_{y})^2} \\ 
% 	& + \frac{\sqrt{[c^i_{1} (\alpha^i_{y})^2 + c^i_{2} (\alpha^i_{x})^2]^2 - 4 (c^i_{3})^2 (\alpha^i_{x})^2 (\alpha^i_{y})^2}}{2 (\alpha^i_{x})^2 (\alpha^i_{y})^2} \notag  \\
% 	& =: \mathrm{C}(\rvec{x}^{i}_t, \alpha^{i}_t) \ , 
% \end{align}
% where 
% \begin{align}\label{eq:covariance_matrix}
%   \mathrm{Cov}(\rvec{x}^i_t) := \begin{bmatrix}
%                                  c^i_{1} & c^i_{3} \\
% 				 c^i_{3} & c^i_{2}
%                                 \end{bmatrix}
% \end{align}
% is the covariance matrix of the position estimate of agent $i$. 
% Since the Chebychev bound for $P^i_t$ will play an important role we define the upper bound on $P^i_t$ in~(\ref{eq:chebychev}) as $\mathrm{C}(\rvec{x}^i_t, \alpha^i_t)$. 
% This inequality holds for position estimates with arbitrary probability distribution. 
The theorem not only provides that the probability that the true position is outside of the region becomes smaller as the size of the region increases, it also gives the constructive rate~(\ref{eq:chebychev}) at which the probability decreases. 
% The theorem gives a precise rate of the decay of the probability that the random vector $\rvec{x}^i_t$ takes values outside of $E^i_t$. 
We will use this upper bound~(\ref{eq:chebychev}) on the rate how $P^i $ decreases to derive a method on how to determine a size of the region $E^i $ such that just the right amount of probability mass lies outside the region. 
Since the bound will play such an important role, we will denote it by $ \mathrm{C} (\rvec{x}^i , E^i )$ and so
\begin{equation}\label{eq:cheb_bound}
  \begin{aligned}
    \mathrm{C} (\rvec{x}^i , E^i ) := \frac{C_{11} (\alpha^i_{y})^2 +  C_{22} (\alpha^i_{x})^2}{2 (\alpha^i_{x})^2 (\alpha^i_{y})^2} 
	       + \frac{\sqrt{[C_{11} (\alpha^i_{y})^2 + C_{22} (\alpha^i_{x})^2]^2 - 4 C_{12}^2 (\alpha^i_{x})^2 (\alpha^i_{y})^2}}{2 (\alpha^i_{x})^2 (\alpha^i_{y})^2} \ .
  \end{aligned}
\end{equation} 
Please note that the bound $\mathrm{C}(\rvec{x}^i , E^i )$ depends on the covariance matrix $(C^i_{kl})_{k,l = 1,2}$ of the uncertain state $\rvec{x}^i $ and the size of the region $E^i $ which is determined by its lateral lengths $\alpha^i_x$ and $\alpha^i_y$. 
% It is therefore possible to control the probability $P^i $ directly via the bound $\mathrm{C}(\rvec{x}^i , \alpha^i )$.  
% So if we would be given a bound 

In the rest of this section we want to investigate the following question: Given an amount $\gamma^i$ of probability mass, how do we determine the size of the region such that at most $\gamma^i$ of the probability mass of the uncertain state $\rvec{x}^i $ lies outside of the region?
Or more formally: \emph{ Given $ 0 \le \gamma^i \le 1$ and an uncertain state $\rvec{x}^i$, how do we determine $\alpha^i_x$ and $\alpha^i_y$ such that
\begin{align}\label{eq:ripp_level}
  \Pr(\rvec{x}^i  \notin E^i  ) \le \gamma^i \ ?
\end{align} } 

For now we assume that we are given some level $0 \le \gamma^i \le 1$ for which we want to determine the size $E^i $ such that~(\ref{eq:ripp_level}) holds for the uncertain state $\rvec{x}^i $. 
The shape of the region $E^i $ is already defined to be rectangular, while its size is determined by the parameters $\alpha^i_{x}$, which is the region's extent in the direction of the $x$ coordinate, and $\alpha^i_{y}$, which is the extent in the $y$ direction.
%
% From the Whittle's Chebychev inequality in Thm.~\ref{thm:theorem1} and the bound~(\ref{eq:chebychev}) with the definition~(\ref{eq:cheb_bound}) 
If we find $\alpha^i_{x}$ and $\alpha^i_{y}$ so that
\begin{align}\label{eq:tmp1}
  \mathrm{C}(\rvec{x}^i , E^i ) = \gamma^i
\end{align}
holds, then by Whittle's Chebychev inequality~(\ref{eq:chebychev}) in Thm.~\ref{thm:theorem1} we can guarantee that 
\begin{align}\label{eq:tmp2}
 \Pr(\rvec{x}^i  \notin E^i ) \le \mathrm{C}(\rvec{x}^i , E^i ) = \gamma^i
\end{align}
the probability mass of the uncertain position $\rvec{x}^i $ outside of the region $E^i $ is at most $\gamma^i$. 
The inequality~(\ref{eq:tmp2}) follows from Whittle's Chebychev inequality~(\ref{eq:chebychev}) in Thm.~\ref{thm:theorem1} together with the definition of $\mathrm{C}(\rvec{x}^i, E^i)$ in~(\ref{eq:cheb_bound}) and Eq.~(\ref{eq:tmp1}). 

So in order to bound the probability mass of the uncertain position of agent $i$ outside of the region $E^i $ we have to determine the parameters $\alpha^i_{x}$ and $\alpha^i_{y}$ such that the equality $\mathrm{C}(\rvec{x}^i , E^i ) = \gamma^i$ holds. 
To achieve this, we have to add another equation, since the equation $\mathrm{C}(\rvec{x}^i , E^i ) = \gamma^i$ is only one equation for the two unknowns $\alpha^{i}_{x}$ and $\alpha^i_{y}$ and hence it is under determined. 
We propose to choose the parameters $\alpha^{i}_{x}$ and $\alpha^i_{y}$ such that additionally
\begin{align}\label{eq:alpha_ratio}
  \frac{\alpha^{i}_{x}}{ \alpha^{i}_{y}} = \sqrt{\frac{C^{i}_{11}}{C^{i}_{22}}} %\  \text{ for } \ \Cov(\rvec{x}^{i} ) = \begin{bmatrix} C^{i}_{11} & C^{i}_{12} \\ C^{i}_{12} &  C^{i}_{22} \end{bmatrix}
\end{align}
holds. 
This choice is motivated by the intuition that for an uncertain position with axis-aligned Gaussian distribution (i.e., $C^{i}_{12} = 0$ in the covariance), the diagonal of the covariance matrix  given through $C^{i}_{11}$ and $C^{i}_{22}$, quantifies the extent of the covariance ellipsoid in $x$-direction and $y$-direction. 
% Since $\alpha^{1}_{x}$ describes the size of the box $E^{1} $ in $x$-direction and $\alpha^{1}_{y}$ in $y$-direction, 
If the ratio of $\alpha^{i}_{x}$ and $\alpha^{i}_{y}$ equals the ratio of $ \scriptstyle \sqrt{C^{i}_{11}}$ and $\scriptstyle \sqrt{C^{i}_{22}}$, the shape and extent of the region $E^i $ follows the shape and extent of the covariance ellipsoid. 
So if there is considerable uncertainty in one of the coordinate directions, indicated by a covariance ellipsoid with strong extent in this direction, the region will also have a stronger spread in this direction to account for this increased uncertainty. 
Other relations like $\alpha^{i}_{x} = \alpha^{i}_{y}$ could be employed, too, and their influence on the region $E^i $ and the resulting constraints will be subject of future research. 

When we insert the equation for $\alpha^i_y$ that results from~(\ref{eq:alpha_ratio})
\begin{align}
  \alpha^i_y = \sqrt{\frac{C^{i}_{22}}{C^{i}_{11}}} \alpha^i_x
\end{align}
into the equation for the bound $C(\rvec{x}^i, E^i) = \gamma^i$ the latter is an equation with only one remaining unknown, namely $\alpha^i_x$. 
% Inserting (\ref{eq:alpha_ratio}) into the Chebychev bound~(\ref{eq:cheb_bound}) yields an equation for the single unknown $\alpha^{i}_{x}$. 
The equation for the remaining unknown is a polynomial of degree four with two real solutions that can be determined analytically. %, with two negative solutions that we discard. 
The positive real solution of the equation is 
% \small
\begin{align}\label{eq:alpha_1}
 \alpha^{i}_{x} = \sqrt{\frac{C^{i}_{11}}{\gamma^i}   + \frac{\sqrt{C^{i}_{11} C^{i}_{22} (C^{i}_{11} C^{i}_{22} - (C^{i}_{12})^2)(\gamma^i)^2  }}{C^{i}_{22} (\gamma^i )^2} } \ .
\end{align}
% {\bf TODO: Diese Gleichung nochmal durch Mathematica jagen, ob sie stimmt!!! DONE}

Given an uncertain position $\rvec{x}^i$ together with the covariance matrix of this uncertain position and given a level $\gamma^i$, Eqs.~(\ref{eq:alpha_1}) and~(\ref{eq:alpha_ratio}) allow us to determine the RIPP region $E^i$ such that $C(\rvec{x}^i,E^i) = \gamma^i$ holds. 
Together with Whittle's Chebychev inequality we can then guarantee that the probability mass of $\rvec{x}^i$ outside the RIPP region $E^i$ is at most $\gamma^i$ and i.e. $\Pr(\rvec{x}^i \notin E^i) \le \gamma^i$!
% \end{align}
So for any given level $\gamma^i$ the construction above enables us to determine a region of increased probability of presence for $\rvec{x}^i$ such that we can precisely control that at most $\gamma^i$ probability mass of $\rvec{x}^i$ lies outside of the region.

\subsection{Collision Avoidance Based on Non-overlapping RIPPs}\label{sec:good_constraints}
In this section we will make use of the results of the previous section to derive our RIPP formulation of collision avoidance constraints. 
We consider two agents $1$ and $2$ with uncertain positions $\rvec{x}^1_t$ and $\rvec{x}^2_t$ together with an upper bound $0 \le \delta^{1,2}_t \le 1$ on the probability of a collision of these agents at time instance $t$. 
We will derive the RIPP formulation of collision avoidance constraints and then prove in Theorem~\ref{thm:theorem2} that controls found with the novel RIPP constraints are feasible for the original problem with bound $\delta^{1,2}_t$ on the probability of a collision between agents $1$ and $2$. 

The RIPP constraints are constructed in two steps: 
% In the first step we split the upper bound $\delta^{1,2}_t$ into two positive parts $\gamma^1_t$ and $\gamma^2_t$ such that $\delta^{1,2}_t = \gamma^1_t  + \gamma^2_t$.
In the first step we determine the RIPPs $E^1_t$ and $E^2_t$ for agents $1$ and $2$ such that the probability mass outside of the RIPPs is at most some $\gamma^i_t$ for both $i=1,2$. % e respective $\gamma^i_t$s.  $C(\rvec{x}^1_t, E^1_t) = \gamma^1_t$ and $C(\rvec{x}^2_t, E^2_t) = \gamma^2_t$, ie. the probability mass outside of the RIPPs is bounded by the respective $\gamma^i_t$s. 
% This means that not more than $\gamma^i_t$ probability mass lies outside of the RIPP for agent $i$.  
% The bound $\delta^{1,2}_t$ can be understood as a bound on the maximum joint probability mass of $\rvec{x}^1_t$ and $\rvec{x}^2_t$ that is allowed to come closer than the minimum distance between the agents. 
% By splitting it into two parts and using each of these parts to construct a RIPP region we decompose the joint probability of a collisions and distribute parts of it onto each agent. 
In the second step a constraint is added to the optimization problem that warrants that the two RIPPs will not overlap at time step $t$. 
% In Theorem~\ref{thm:theorem2} the main result of this section we prove that controls found under these novel RIPP constraints are feasible for the original problem with chance constraints on the probability of inter-agent collisions.  
% In short, the three steps are
% \begin{enumerate}
%   \item The agents split the upper bound on the probability of a collision into two positive parts $\gamma^1_t$ and $\gamma^2_t$ such that $\delta^{1,2}_t = \gamma^1_t  + \gamma^2_t$.
%   \item They determine their respective RIPPs $E^1_t$ and $E^2_t$ such that $C(\rvec{x}^1_t, E^1_t) = \gamma^1_t$ and $C(\rvec{x}^2_t, E^2_t) = \gamma^2_t$. 
%   \item A constraint $\mathcal{C}_t$ is introduced that warrants that the RIPPs $E^1_t$ and $E^2_t$ do not overlap at time step $t$. 
% \end{enumerate}
% 
% We propose that agents $1$ and $2$ split the chance constraint bound $\delta^{1,2}_t$ into two parts
% \begin{align}\label{eq:delta_eq}
%  \gamma^1_t = \frac{1}{d}\delta^{1, 2}_t \text{ and } \gamma^2_t = \frac{d-1}{d}\delta^{1,2}_t \ ,
% \end{align}
% with free parameter $d > 1$. 
% In our simulations in Section~\ref{sec:sim}, we used an even split at $d=2$. 
% One could also employ some kind of negotiation algorithm to find a split that is optimal for the agents and this will be subject of future work. 
 
% To determine the RIPP regions  $\mathrm{C}(\rvec{x}^{1}_tagents $1$ and $2$ have to conduct the following two steps: 
The construction of the RIPP regions involves the mean of the uncertain states, since those determine the position of the RIPP, it involves the covariance of the uncertain states, since the computation of the Chebychev bound~(\ref{eq:chebychev}) involves the covariance matrix, and finally the parameters $\alpha^{i}_x$ and $\alpha^{i}_y$ determine the size of the RIPP region. 
% \begin{itemize}
%   \item Determine the covariance of their position estimate at time instance $t$ since the bound~(\ref{eq:chebychev}) involves it,
%   \item Determine the size parameters $\alpha^{1}_t$ and $\alpha^{2}_t$ for the boxes $E^{2}_t$ and $E^{2}_t$.
% \end{itemize}
% The RIPP for each agent is centered at the mean of its uncertain state and its size is determined by the covariance of uncertain state. 

We assume without loss of generality that the system noise $\rvec{\nu}^i_t$ for all agents and for all time steps is zero-mean. 
(If the noise had a non-zero mean, we could subtract this mean from the system equations as deterministic disturbance and would have reduced this noise to zero-mean noise again.) 
For system noise with vanishing mean the mean $\vec{\mu}^i_t$ of agent $i$'s uncertain state follows the recursive rule
\begin{align}
  \vec{\mu}^i_{t} = \vec{\mu}^i_{t-1} + \mat{B} \vec{u}^i_{t-1} \ ,
\end{align}
as one can easily check. 
% or
% \begin{align}
%   \vec{\mu}^i_t = \vec{\mu}^i_0 + \sum_{k = 0}^t \mat{B} \vec{u}^i_k \ ,
% \end{align}
% where $\vec{\mu}^i_0$ is the mean of the initial uncertain state $\rvec{x}^i_0$. 
% These formulas allow us to determine the mean of the uncertain state of any agent at any time step. 
Please note that the mean always \emph{depends linearly} on the control inputs. 

The covariances of the uncertain states $\rvec{x}^i_t$ do not depend on any control inputs, but only on the covariances of the prior distribution $\rvec{x}^i_0$ and the noise terms $\rvec{\nu}^i_{1:t}$. 
% For the case of additive noise with general distribution this follows from the fact that for a multivariate random variable $\rvec{X}$ and a deterministic vector $\vec{b}$ of same dimension $\mathrm{Cov}(\rvec{X} + \vec{b}) = \mathrm{Cov}(\rvec{X})$ holds. 
The recursive formula for the evolution of the covariances is
\begin{align}
  \mathrm{Cov}(\rvec{x}^i_{t}) & = \mat{A}^i \mathrm{Cov}(\rvec{x}^i_{t-1}) (\mat{A}^i)^T + \mathrm{Cov}(\rvec{\nu}^i_{t-1}) \label{eq:rec_cov} \\ 
				 & + \mat{A}^i\mathrm{Cov}(\rvec{x}^i_{t-1}, \rvec{\nu}^i_{t-1}) + \mathrm{Cov}(\rvec{\nu}^i_{t-1}, \rvec{x}^i_{t-1})(\mat{A}^i)^T \notag \ . 
\end{align}
This property can be derived from basic matrix manipulations and covariance matrix properties. 
In Section~\ref{sec:gen_prob}, we assumed that covariances of the prior distributions $\rvec{x}^i_0$ and the noise terms $\rvec{\nu}^i_t$ are known in advance, so the agents can recursively compute the covariance of their uncertain state at time step $t$. % $\Cov(\rvec{x}^i_t)$. 

The chance constraint bound $\delta^{1,2}_t$ can be understood as a bound on the maximum joint probability mass of $\rvec{x}^1_t$ and $\rvec{x}^2_t$ that is allowed to come closer than a minimum distance. % between the agents. 
We will split this upper bound on the joint probability into two parts $\gamma^1_t$ and $\gamma^2_t$ and distribute these parts onto the agents to construct a RIPP region for each agent. 
% The RIPP region is determined such that $C(\rvec{x}^1_t, E^1_t) = \gamma^1_t$ and $C(\rvec{x}^2_t, E^2_t) = \gamma^2_t$.
We propose that agents $1$ and $2$ split the chance constraint bound $\delta^{1,2}_t$ into parts according to
\begin{align}\label{eq:delta_eq}
 \gamma^1_t = \frac{1}{d}\delta^{1, 2}_t \text{ and } \gamma^2_t = \frac{d-1}{d}\delta^{1,2}_t \ ,
\end{align}
with free parameter $d > 1$. 
 % we decompose the and distribute parts of it onto each agent. 
In our simulations in Section~\ref{sec:sim}, we used an even split at $d=2$. 
One could also employ a negotiation algorithm to find a split that is optimal for the agents and this will be subject of future work. 
Equations $C(\rvec{x}^1_t, E^1_t) = \gamma^1_t$ and $C(\rvec{x}^2_t, E^2_t) = \gamma^2_t$ then uniquely determine the $\alpha$-size parameters for the RIPP regions. 
% We propose that the size of the RIPPs is determined such that $C(\rvec{x}^i_t, E^i_t) = \gamma^i_t$ for $i=1$ and $i=2$.  

Now we have everything we need to construct the RIPP regions for agents $1$ and $2$, the means and covariances of their uncertain states and a size for the RIPPs. 
Denote by $E^1_t$ and $E^2_t$ the RIPPs constructed from these parameters. 

Finally, all we have left to do is to define the constraint that ensures that these RIPPs do not overlap:
\begin{definition}[Constraint $\mathcal{C}_t$]
The expected values $\vec{\mu}^{1}_t$ and $\vec{\mu}^{2}_t$ have a distance of more than $\frac{1}{2} (\alpha^{1}_{t,x} + \alpha^{2}_{t,x}) + \epsilon$ in the $x$-direction, i.e. $\vert \mu^{1}_{t,x} - \mu^{2}_{t,x} \vert > \frac{1}{2} (\alpha^{1}_{t,x} + \alpha^{2}_{t,x}) + \epsilon$ \emph{or} a distance of more than $\frac{1}{2} (\alpha^{1}_{t,y} + \alpha^{2}_{t,y}) + \epsilon$ in the $y$-direction, i.e. $\vert \mu^{1}_{t,y} - \mu^{2}_{t,y} \vert > \frac{1}{2} (\alpha^{1}_{t,y} + \alpha^{2}_{t,y}) + \epsilon$.
\end{definition}
Recall that the parameters $\alpha^i_{t,x}$ and $\alpha^i_{t,y}$ in this definition are directly related to the lateral lengths of the RIPP $E^i_t$. 
Constraint $\mathcal{C}_t$ is then a simple constraint on the means of the uncertain states of agents $1$ and $2$ depending on the sizes of the RIPP regions. 

The following theorem proves that controls for agents $1$ and $2$ for that constraint $\mathcal{C}_t$ holds, satisfy that the probability of a collision between agent $1$ and $2$ at time step $t$ is less or equal than the bound $\delta^{1,2}_t$, i.e. they are feasible for the problem with bounds on inter-agent collision probabilities. 
\begin{thm}\label{thm:theorem2}
  Let $F^c$ be the feasible region for inter-agent collision avoidance and $\rvec{x}^{1}_t$, $\rvec{x}^{2}_t$ the uncertain positions of agents $1$ and $2$ at time step $t$. 
  Further let $\gamma^1_t$ and $\gamma^2_t$ be as above such that $\gamma^1_t + \gamma^2_t = \delta^{1,2}_t$ and let $E^1_t$ and $E^2_t$ be RIPPs so that $C(\rvec{x}^i_t, E^i_t) = \gamma^i_t$. 
  Then for any control sequences $\vec{u}^1_{1:H}$ for agent $1$ and $\vec{u}^2_{1:H}$ for agent $2$ for that constraint $\mathcal{C}_t$ is satisfied at time step $t \le H$, the probability of a collision of agents $1$ and $2$ is below $\delta^{1,2}_t$
    \begin{align}
      \Pr((\rvec{x}^{1}_t, \rvec{x}^{2}_t) \notin F^c) \le \delta^{1,2}_t \ .
    \end{align}
\end{thm}
It follows that if the control sequences also satisfy the other constraints in the formulation of the original problem in Section~\ref{sec:gen_prob} they are feasible for the MPC problem with chance constraints on the probability of inter-agent collisions! 

In order to prove Theorem~\ref{thm:theorem2}, we will first establish the following Lemma.

%%%%%%%%%%%%%%%%%%%%%%%%%%%%%%%%%%%%%%%%%%%%%%LEMMA%%%%%%%%%%%%%%%%%%%%%%%%%%%%%%%%%%%%%%%%%%%%%%%%%%%%%%%%%%%%%%%%%%%%%%%%%%%%%%%%%%%%%%%%%%%%%%%%%%%%%%%%%%%%%%%%%%%%%
\begin{lemma}\label{lemma:lemma}
  Let $F^c$, $\rvec{x}^{1}_t$, $\rvec{x}^{2}_t$, $P^{1}_t$ and $P^{2}_t$ be as above. 
  If constraint $\mathcal{C}_t$ is satisfied, the inequality
  \begin{align}\label{eq:tighter_bound}
    \Pr((\rvec{x}^{1}_t , \rvec{x}^{2}_t) \notin F^c ) \le P^{1}_t + P^{2}_t
  \end{align}
 holds, where as in the previous section $P^i_t = \Pr(\rvec{x}^i_t \notin E^i_t)$.   
\end{lemma}
%%%%%%%%%%%%%%%%%%%%%%%%%%%%%%%%%%%%%%%%PROOF%%%%%%%%%%%%%%%%%%%%%%%%%%%%%%%%%%%%%%%%%%%%%%%%%%%%%%%%%%%%%%%%%%%%%%%%%%%%%%%%%%%%%%%%%%%%%%%%%%%%%%%%%%%%%%%%%%%%%%%%%%%%
\proof 
By marginalization we have
% \small
\begin{equation} \label{eq:tot_prob}
  \begin{aligned}
    \Pr  & ((\rvec{x}^{1}_t , \rvec{x}^{2}_t )  \notin  F^c ) \\ 
							     &= \Pr((\rvec{x}^{1}_t , \rvec{x}^{2}_t) \notin F^c, \rvec{x}^{1}_t \in E^{1}_t, \rvec{x}^{2}_t \in E^{2}_t ) \\ 
							     & \quad  + \Pr((\rvec{x}^{1}_t , \rvec{x}^{2}_t) \notin F^c, \rvec{x}^{1}_t \in E^{1}_t, \rvec{x}^{2}_t \notin E^{2}_t ) \\
							     & \quad + \Pr((\rvec{x}^{1}_t , \rvec{x}^{2}_t) \notin F^c, \rvec{x}^{1}_t \notin E^{1}_t, \rvec{x}^{2}_t \in E^{2}_t ) \\ 
							     & \quad + \Pr((\rvec{x}^{1}_t , \rvec{x}^{2}_t) \notin F^c, \rvec{x}^{1}_t \notin E^{1}_t, \rvec{x}^{2}_t \notin E^{2}_t ) \ .
  \end{aligned}
\end{equation}
We will first show that the probability
\begin{align}\label{eq:zero_prob}
  \Pr ((\rvec{x}^{1}_t , \rvec{x}^{2}_t) \notin F^c,\rvec{x}^{1}_t \in E^{1}_t,\rvec{x}^{2}_t \in E^{2}_t ) 
%       & = \int_{E^{1}_t} \int_{E^{2}_t} \chi_{CF^c}(\vec{x}^{1}_t, \vec{x}^{2}_t) f(\vec{x}^{1}_t) f(\vec{x}^{2}_t) d \vec{x}^{1}_t d \vec{x}^{2}_t \ .
\end{align}
is zero by showing that $(E^{1}_t \times E^{2}_t) \cap F^c = \emptyset$ from which the claim follows because then the event $\rvec{x}^{1}_t \in E^{1}_t \wedge \rvec{x}^{2}_t \in E^{2}_t \wedge (\rvec{x}^{1}_t , \rvec{x}^{2}_t) \in F^c$ has zero probability mass. 
We have $F^c \subset F^c_{\infty} := \{ (\vec{x},\vec{y}) \vert \ \Vert \vec{x} - \vec{y} \Vert_{\infty} < \epsilon\}$ because of the the norm inequality $\Vert \vec{x} \Vert_{\infty} \le \Vert \vec{x} \Vert_2$. 
Hence, $(E^{1}_t \times E^{2}_t) \cap F^c \subset (E^{1}_t \times E^{2}_t) \cap F^c_{\infty}$ and we will show $(E^{1}_t \times E^{2}_t) \cap F^c_{\infty} = \emptyset$, then $(E^{1}_t \times E^{2}_t) \cap F^c = \emptyset$ and the probability~(\ref{eq:zero_prob}) will be zero. 
Let $\vec{x}^{1}_t \in E^{1}_t$ and $\vec{x}^{2}_t \in E^{2}_t$, and for contradiction assume that $(\vec{x}^{1}_t, \vec{x}^{2}_t) \in F^c_{\infty}$. 
For the $x$-coordinates of $\vec{x}^{1}_t$ and $\vec{x}^{2}_t$ we have that $\vert x^{1}_{t,x} - x^{2}_{t,x} \vert < \epsilon$ and also for the $y$-coordinates $\vert x^{1}_{t,y} - x^{2}_{t,y} \vert < \epsilon$ since by the definition of $F^c_{\infty}$ it holds that $\Vert \vec{x}^{1}_t - \vec{x}^{2}_t \Vert_{\infty} = \max \{ \vert x^{1}_{t,x} - x^{2}_{t,x} \vert, \vert x^{1}_{t,y} - x^{2}_{t,y} \vert \} < \epsilon$. 
For the expected values we have $\vert \mu^{1}_{t,x} - \mu^{2}_{t,x} \vert \le \vert \mu^{1}_{t,x} - x^{1}_{t,x} \vert + \vert x^{1}_{t,x} - x^{2}_{t,x} \vert  + \vert x^{2}_{t,x} - \mu^{2}_{t,x} \vert < \epsilon + \frac{1}{2}\alpha^{1}_{t,x} + \frac{1}{2} \alpha^{2}_{t,x}$ because $\vert x^{1}_{t,x} - \mu^{1}_{t,x} \vert \le \frac{1}{2}\alpha^{1}_{t,x}$ and $\vert x^{2}_{t,x} - \mu^{2}_{t,x} \vert \le \frac{1}{2}\alpha^{2}_{t,x}$ per definition of the boxes $E^{1}_t$ and $E^{2}_t$. 
The same is true for the $y$-coordinate and, thus, we have constructed a contradiction to the assumption that constraint $\mathcal{C}_t$ holds. 
%
% Let $\vec{x}^{1}_t \in E^{1}_t$ and $\vec{x}^{2}_t$, assume that 
%%%%%%%%%%%%%%%%% End
Thus, $\Pr((\rvec{x}^{1}_t , \rvec{x}^{2}_t) \notin F^c, \rvec{x}^{1}_t \in E^{1}_t,\rvec{x}^{2}_t \in E^{2}_t ) = 0$ holds if constraint $\mathcal{C}_t$ is satisfied.

For the second summand in (\ref{eq:tot_prob}), we have %$\Pr((\rvec{x}^{1}_t , \rvec{x}^{2}_t) \notin F^c, \rvec{x}^{1}_t \in E^{1}_t, \rvec{x}^{2}_t \notin E^{2}_t )$ can be bounded as follows
\begin{align}
  \Pr((\rvec{x}^{1}_t & , \rvec{x}^{2}_t) \notin F^c, \rvec{x}^{1}_t \in E^{1}_t, \rvec{x}^{2}_t \notin E^{2}_t ) \\  
		      & \le \Pr(\rvec{x}^{1}_t \in E^{1}_t, \rvec{x}^{2}_t \notin E^{2}_t ) \\
		      & = \Pr(\rvec{x}^{1}_t \in E^{1}_t) \Pr(\rvec{x}^{2}_t \notin E^{2}_t ) = (1 -P^{1}_t) P^{2}_t \ , \notag
\end{align}
where the inequality follows from the fact that intersecting with an additional event can only decrease their probability mass. 
The first equality follows from the assumed independence of the position estimates of agents $1$ and $2$. 

The third and fourth summand in (\ref{eq:tot_prob}) can be bounded with the same arguments as the second summand, with the fourth summand being bounded by $P^{1}_t P^{2}_t$. 
Through summation of all the bounds, we obtain
\begin{align}
  \Pr & ((\rvec{x}^{1}_t , \rvec{x}^{2}_t) \notin F^c ) \\
      & \le (1 -P^{1}_t) P^{2}_t + (1 -P^{2}_t) P^{1}_t + P^{1}_t P^{2}_t \\
      & = P^{1}_t + P^{2}_t - P^{1}_t P^{2}_t \ .
\end{align}

Since both $P^{1}_t \ge 0$ and $P^{2}_t \ge 0$ hold and then also $P^{1}_t P^{2}_t \ge 0$, we have the slightly more coarse inequality
\begin{align}
  \Pr((\rvec{x}^{1}_t , \rvec{x}^{2}_t) \notin F^c) \le P^{1}_t + P^{2}_t \ . \quad \quad \blacksquare
\end{align}
% $\blacksquare$

We have thus shown that the probability of a collision $\Pr((\rvec{x}^{1}_t, \rvec{x}^{2}_t) \notin F^c)$ can be bounded from above by the probabilities that the uncertain states $\rvec{x}^{1}_t$ and $\rvec{x}^{2}_t$ are outside the RIPP regions $E^1_t$ and $E^2_t$. %take values outside boxes around their expected values, if the distance of these expected values is large enough. 
% Before we have shown in Theorem~\ref{thm:theorem1} how these probabilities decrease as the sizes of the boxes increase. 
Summarizing the results of Lemma~\ref{lemma:lemma} and the construction of the RIPPs together with Theorem~\ref{thm:theorem1} we can deduce Theorem~\ref{thm:theorem2}:
\begin{align}
  \Pr((\rvec{x}^{1}_t, \rvec{x}^{2}_t) \notin F^c) & \le P^{1}_t + P^{2}_t \\
						     & \le \mathrm{C}(\rvec{x}^{1}_t, \alpha^{1}_t) + \mathrm{C}(\rvec{x}^{2}_t, \alpha^{2}_t) = \gamma^1_t + \gamma^2_t \\
						     & = \delta^{1,2}_t
\end{align}
if constraint $\mathcal{C}_t$ holds, where $\mathrm{C}(\rvec{x}^{1}_t, \alpha^{1}_t)$ and $ \mathrm{C}(\rvec{x}^{2}_t, \alpha^{2}_t)$ are the Chebychev bounds as defined in (\ref{eq:chebychev}). 

The theorem warrants that if we solve the chance constrained MPC problem with RIPP constraints of the form $\mathcal{C}_t$, then the obtained controls are automatically feasible for the chance constrained MPC problem with full constraints on inter-agent collision probabilities. 
Hence, the novel RIPP constraints allow us to find solutions to the MPC problem with the complicated probabilistic coupling constraints without any knowledge about the agents' uncertain states besides the covariance and also without ever computing or evaluating the probability of a collision of two agents. 

It suggests itself that controls found for the MPC problem with RIPP constraints are slightly more suboptimal than controls for the MPC problem with full probabilistic constraints, since the RIPP constraints are a conservative tightening. 
While this is true, we will provide empirical evidence in our simulations that the degree of subtoptimality is very low. 

\subsection{Constraint Formulation for RIPP Method}

In this section we will outline how the RIPP constraints can be formulated as mixed integer constraints.
In summary, the RIPP constraints for a time instance $t$ can be constructed as follows:
\begin{itemize}
  \item For all agents $i = 1, \dots, M$ we determine $\Cov(\rvec{x}^i_t)$ according to~(\ref{eq:rec_cov}). Please note, that this is not done at run time of the mixed integer linear optimization routine but before this routine is started. 
  \item Also before the mixed integer solver starts, for all agent combinations $1$ and $2$ and upper bounds on the collision probability $\delta^{1,2}_t$ we determine the RIPP regions such that $\mathrm{C}(\rvec{x}^{1}_t, E^{1}_t)$ and $\mathrm{C}(\rvec{x}^{2}_t, E^{2}_t)$ equal $\frac{1}{d} \delta^{1,2}_t$ and $\frac{d-1}{d} \delta^{1,2}_t$ respectively. 
  \item Then we replace the probabilistic collision avoidance constraint~(\ref{eq:ca_constraint}) by the mixed integer linear formulation of constraint $\mathcal{C}_t$. 
  \item Then the mixed integer linear optimization routine solves the MILP.
\end{itemize}

Constraint $\mathcal{C}_t$ for agents $1$ and $2$ is that at least one of the coordinates in the 2D-plane of their expected values have a distance of at least $\frac{1}{2}(\alpha^{1}_{t,x} + \alpha^{2}_{t,x}) + \epsilon$ or $\frac{1}{2}(\alpha^{1}_{t,y} + \alpha^{2}_{t,y}) + \epsilon$ respectively. 
We have already highlighted in the last section how the covariances of the uncertain states of the agents can be determined and how the size of the RIPP regions is obtained. 
The position of the RIPP region depends on the mean of an agent's uncertain positions. 
Since this mean depends on the applied control inputs, it becomes a decision variable and it is determined through
\begin{align}\label{eq:mean_con}
  \vec{\mu}^i_t = \vec{\mu}^i_{t-1} + \sum_{k=1}^{t -1} \mat{B} \vec{u}^i_k
\end{align}
for each agent $i$. 
% Since the expected values of the state distributions are not known, the agents have to approximate them by the sample means that converge to the true means for $N \to \infty$. 
% So we introduce variables 
% \begin{align}
% \mu^{1}_{t} = \frac{1}{N} \sum_{j = 1}^{N}  \vec{x}^{1}_{t,j} \text{ and }  \mu^{2}_{t} = \frac{1}{N}  \sum_{l = 1}^{N} \vec{x}^{2}_{t,l} \ , 
% \end{align}
% the two sample means of the position estimates of the agents through linear constraints.
% \begin{align}
%  & \vec{\mu}^{1}_{t} = \sum_{j = 1}^{N} w^{1}_{t,j} \vec{x}^{1}_{t,j} \\ & 
%    \vec{\mu}^{2}_{t} = \sum_{l = 1}^{N} w^{2}_{t,l} \vec{x}^{2}_{t,l} \ .
% \end{align}
Now we model the RIPP constraint $\mathcal{C}_t$ for a time step $t$ in the planning horizon and a pair of agents denoted by $1$ and $2$ as constraints for the mixed integer optimization routine. 
Therefor, we introduce constraints on the distance of the means to model the RIPP constraint $\mathcal{C}_t$ 
% \small
\begin{align}
 & \mu^{1}_{t,x} - \mu^{2}_{t,x} \ge \frac{1}{2}(\alpha^{1}_{t,1} + \alpha^{2}_{t,1}) + \epsilon \text{ or } \\ &
   \mu^{2}_{t,x} - \mu^{1}_{t,x} \ge \frac{1}{2}(\alpha^{1}_{t,1} + \alpha^{2}_{t,1}) + \epsilon \text{ or } \\ &
   \mu^{1}_{t,y} - \mu^{2}_{t,y} \ge \frac{1}{2}(\alpha^{1}_{t,2} + \alpha^{2}_{t,2}) + \epsilon \text{ or } \\ &
   \mu^{2}_{t,y} - \mu^{1}_{t,y} \ge \frac{1}{2}(\alpha^{1}_{t,2} + \alpha^{2}_{t,2}) + \epsilon \ , 
\end{align}
% \normalsize
where again the subscripts $x$ and $y$ denote the components of the means in the $x$ and $y$ axis. 
Since a mixed integer linear solver can not directly understand these logical ``or``-constraints, we formulate them as logical ``and``-constraints with the ''Big M''-method as in Sec.~\ref{sec:bad_constraints}
\begin{align}
& \mu^{1}_{t,x} - \mu^{2}_{t,x} \ge \frac{1}{2} ( \alpha^{1}_{t,1} + \alpha^{2}_{t,1}) + \epsilon - M_o b^1_{t,1,2} \text{ and } \label{eq:mean_con2} \\ &
  \mu^{2}_{t,x} - \mu^{1}_{t,x} \ge \frac{1}{2} ( \alpha^{1}_{t,1} + \alpha^{2}_{t,1}) + \epsilon - M_o b^2_{t,1,2} \text{ and } \\ &
  \mu^{1}_{t,y} - \mu^{2}_{t,y} \ge \frac{1}{2} ( \alpha^{1}_{t,2} + \alpha^{2}_{t,2}) + \epsilon - M_o b^3_{t,1,2} \text{ and } \\ &
  \mu^{2}_{t,y} - \mu^{1}_{t,y} \ge \frac{1}{2} ( \alpha^{1}_{t,2} + \alpha^{2}_{t,2}) + \epsilon - M_o b^4_{t,1,2} \text{ and } \\ &
  b^i_{t,1.2} \in \{ 0,1 \} \text{ and } \\ &
  \sum_{i = 1}^4 b^i_{t,1,2} \le 3 \ ,\label{eq:mean_con3}
\end{align}
with large positive number $M_o$. 
Please note, that the proceeding is exactly the same as in Section~\ref{sec:bad_constraints}: if one of the binary variables $b^i_{t,1,2}$ equals one it is possible that the corresponding constraint on the distance of the means is not satisfied. 
Since at least one of the four constraints in the ``or''-formulation above has to be satisfied in order to guarantee that the means are far enough apart, we limit the number of binary variables $b^i_{t,1,2}$ that are allowed to be equal to one by three. 
The equality constraint~(\ref{eq:mean_con}) for the mean and the inequality constraints~(\ref{eq:mean_con2})-~(\ref{eq:mean_con3}) form the mixed integer linear constraints equivalent to the RIPP constraint $\mathcal{C}_t$. 

% \vspace*{-3em}
\section{Complexity Analysis - Number of Binary Constraints}\label{sec:comp_ana}
% \begin{itemize}
In this section we provide a theoretical analysis of the complexity of the MILPs resulting from the sample-based approximation of the probability of inter-agent collisions and from the RIPP constraints. 

In general, solving even reasonably large linear programs can be performed efficiently by standard methods such as the interior points or simplex methods  \cite{Boyd2004,Bertsimas1997}. 
However, solving programs that also include binary or integer variables is generally NP-hard \cite{Bertsimas1997}. 
Recent work like \cite{Schouwenaars2001,Bellingham2002,Schouwenaars2004,Yilmaz2008} has shown that in practice for problems of moderate sizes MILP solvers are efficient enough to employ them under real time requirements. 
The restriction of the manageability of MILPs to problems with only moderately many integer variables is the motivation why we will understand the number of binary variables as a measure of the complexity of the program when comparing the complexity of different formulations of inter-agent collision avoidance constraints. 

% With this definition in mind we analyze the complexity of the approximated collision avoidance formulation detailed above. 
First, we will analyze the number of binary variables the constraints derived from the sample-based approximation of collision probabilities introduce to the program. 
According to the collision avoidance constraint~(\ref{eq:ca_constraint}), each agent has to check for a collision with every other agent at each time instance, resulting in $\frac{M(M-1)}{2} H$ checks for collisions. 
In each of these, the approximated chance constraint~(\ref{eq:pr_crash}) has to be evaluated, resulting in $N^2$ ``sample evaluations''. 
For every ``sample evaluation'', we have a fixed number $b_{cc}$ of binary variables. 
So all in all, we have at least $b_{cc} N^2 \frac{1}{2} M(M-1) H \in \mathcal{O}(N^2 M^2 H)$ binary variables. 
Hence, the number of binary variables of the approximated inter-agent collision chance constraint depends quadratically on the number of agents and quadratically on the number of samples in the state approximations. 
Since the accuracy of the approximation~(\ref{eq:pr_crash}) of the chance constraint depends on the number of samples, a quadratic dependence of the number of binary variables on the number of samples is a major drawback of this formulation. 

Next, we count the number of binary variables the mixed integer linear formulation of the RIPP constraints introduces to the MILP. 
Again, we have to evaluate collision avoidance at every time instance in the planning horizon and for every pairing of agents. 
As a result, the total number of collision avoidance checks remains $H\frac{M(M-1)}{2}$. 
However, each of these checks requires only a fixed number $b_{mb}$ of binary variables that is independent of the number of samples of the state representations.
This can be easily seen from the construction of the RIPP constraints, since the constraints only depend on the means of the uncertain states of the agents and not on any samples!  
The overall number of binary variables for inter-agent collision avoidance is $b_{mb}H\frac{M(M-1)}{2} \in \mathcal{O}(HM^2)$. 

For general multi-robot or multi-UAV systems, the number of agents will typically be in the tens, whereas the number of samples to represent an agent's position estimate will usually be in the hundreds. 
Hence, we achieved a significant reduction of complexity with this approximative formulation of inter-agent collisions. 
% \end{itemize}

% \begin{figure}[t!]
%   \centering
%   \includegraphics[scale=0.4]{fig/alpha_plot_multi_modal_vert }
%   \caption{The image above shows 10000 samples drawn from a bimodal bivariate Gaussian mixture density. 
% % 	  The black dot in the middle depicts the mean of the Gaussian. 
% 	  The parameter $\alpha^i_t$ is calculated according to Eqs.~(\ref{eq:alpha_1}) and (\ref{eq:alpha_ratio}) such that $\mathrm{C}(\rvec{x}^i_t, \alpha^i_t) = 0.05,\ 0.10, \ 0.20$, respectively. 
% 	  The boxes $E^i_t$ corresponding to these bounds as defined in (\ref{eq:def_box}) are depicted in red. 
% 	  The true probability mass of the random variable $\rvec{x}^i_t$ outside of the box $E^i_t$ is estimated by the empiric sample expectation and is $P^i_t = 0.02, \ 0.08, \ 0.13$.}
% % 	  We chose Gaussian distribution for generating the samples just for the convenience, the calculation of the conservative bounds does not rely on this assumption.}
%   \label{fig:boxes}
% %   \vspace*{-2em}
% \end{figure}

\section{Simulations}\label{sec:sim}

% 
% \begin{table}
%   \begin{tabular}{|p{1.2cm}||p{1.2cm}|p{1.2cm}||p{1.2cm}|p{1.2cm}|}
%     \hline 
%     100 MC runs			&	SA, $M=2$ 			& 	DM, $M=2$			&	SA, $M=3$			&	DM, $M=3$	\\ \hline
%     $\frac{\text{mean}}{MNH}$ 	&	316.5279			& 	317.0045			& 	320.0120			& 	331.4733	\\ \hline
% %     $\frac{\sigma}{MNH}$	&	22.7345				&	21.0957				&	14.9088				& 	20.1845		\\ \hline
%   \end{tabular}
%   \caption{Objective values of SA and DM}
% \end{table}
% \begin{figure}[tbh]
%   \centering
%   \includegraphics[scale = 0.4]{fig/runtime_vs_samples_ai }
%   \caption{Run times averaged over 100 Monte Carlo runs with standard deviation for 5 agents, no obstacles and an increasing number of samples for mean based approach. }
% \end{figure}
In our simulations, we consider path planning for multiple UAVs whose movements are affected by wind disturbances. 
We assume that the UAVs all fly at the same fixed height and therefore consider collision avoidance in the two-dimensional plane. 
The task of the UAVs is to reach a certain goal point on a direct path as quickly as possible in order to save fuel. 
Bounds on the control inputs are given through bounds on the maximum acceleration and bounds on the maximum speed at which the UAVs can fly. 
We assume that the same bounds on acceleration and speed apply for all UAVs. 
\subsection{Model Parameters}
We assume that the UAVs all have the same linear motion model given by the double integrator model
\begin{align}
 \rvec{x}^i_t = \mat{A} \rvec{x}^i_{t-1} + \mat{B} \vec{u}^i_{t-1} + \rvec{\nu}^i_{t-1}
\end{align}
with
\begin{align}\label{eq:UAV_dynamics}
  \rvec{x}^i_t = [\rv{x}^i_t, \ \rv{y}^i_t, \ \dot \rv{x}^i_t, \ \dot \rv{y}^i_t]^T \ , \ \vec{u}^i_t = [\ddot x^i_t, \ddot y^i_t]^T \ , \ \Vert \vec{u}^i_t \Vert_\infty \le 12 
\end{align}
and
\begin{equation*}
  \begin{aligned}
      & \mat{A} = \begin{bmatrix}
		    1 & 0 & 1 & 0 \\
		    0 & 1 & 0 & 1 \\
		    0 & 0 & 1 & 0 \\
		    0 & 0 & 0 & 1 \\
		  \end{bmatrix} \ , \ 
      \mat{B} = \begin{bmatrix}
		  0 & 0 \\
		  0 & 0 \\
		  1 & 0 \\
		  0 & 1 \\
		  \end{bmatrix} \ .
\end{aligned}
\end{equation*}
We assume that the initial uncertain states of the agents have Gaussian distribution with the following covariance
\begin{align}
      & \rvec{x}^i_0 \sim \mathcal{N}(\vec{\mu}^i, \mat{C}^i_0) \ , \ \mat{C}^i_0 = \diag[10^{-3}, \ 10^{-3}, \ 10^{-5}, \ 10^{-5}] \ .
\end{align}
We assume that the target way points are given as two-dimensional positions $\vec{Z}^i = [Z^i_1, \ Z^i_2 ]^T$ for each agent and that the objective function is the distance of the position to the target way point (normalized for better comparability with $(HMN)^{-1}$):
\begin{align}
  h^i(\vec{x}^i_t) = (HMN)^{-1} \left\{ \vert x^i_t - Z^i_1\vert + \vert y^i_t - Z^i_2 \vert \right\} \ .
\end{align}
The length of the planning horizon was set to $H=7$ in all simulations.
%       & \vec{u}^i_t \in [-12,12]^2 \ , \ 
%   \end{aligned}
% \end{equation*}

% \begin{itemize}
%   \item 
The disturbance samples $\vec{\nu}^i_{t,j}$ affecting the UAVs are drawn from the discrete Dryden low-altitude model to simulate wind turbulence acting on the UAVs \cite{MILF8785C}. 
%   \item 
The UAVs are assumed to fly with a maximum speed of 45 feet per \second $\ $ at a fixed altitude of 200 feet through a field with light turbulence with wind speed of 15 knots at 20 feet height. 
The minimum distance between the UAVs is set to $\epsilon = 5$~feet. 
The objective of UAV $i$ in all scenarios is to reach a certain random goal way point $\vec{Z}^i$, given as point in the plane, as quickly as possible from a randomly placed starting position given through the mean $\vec{\mu}^i$ of the prior distribution $\rvec{x}^i_0$. % while avoiding other UAVs and obstacles. 
The value of the control objective was divided by $H \cdot M \cdot N$ in all simulations for better comparability of the results.
We randomly placed obstacles of fixed size $50 \times 50$ feet each. 
The mixed integer linear solver we used is CPLEX \cite{Cplex2008}. 

In all simulations we compared the collision avoidance from Sec.~\ref{sec:col_av_app} using the RIPP constraints with the 'full' sample-based approximation of the collision probabilities from Sec.~\ref{sec:col_agents}. 
This is a highly relevant comparison since the sample-based approximations of the collision probabilities are guaranteed to converge to the true collision probabilities when the number of samples approaches infinity and, hence, the control inputs will converge to the optimal control inputs for the chance constrained MPC problem without approximation! 
So the sample-based approximation forms a baseline for any other collision avoidance approximation for the MPC problem with chance constraints on collision probabilities both in the quality of the achieved objective of the control and in run-time.
% \end{itemize}

% \begin{figure}[t]
%   \centering
%   \includegraphics[scale = 0.4]{fig/2agents_2obstacles_sample2_ai }
%   \caption{The plot shows the trajectories of two UAVs computed with the conservative approximation approach (DM). The runtime for this example a standard desktop PC was 33 \second. The runtime for the same scenario computed with the sample approximation approach (SA) took more than 73 \minute. Both formulations solved the MILP to the same objective of 141.6339.}
%   \label{fig:sample}
% \end{figure}

\begin{figure}[htpb]
  \centering
  \includegraphics[scale = 1]{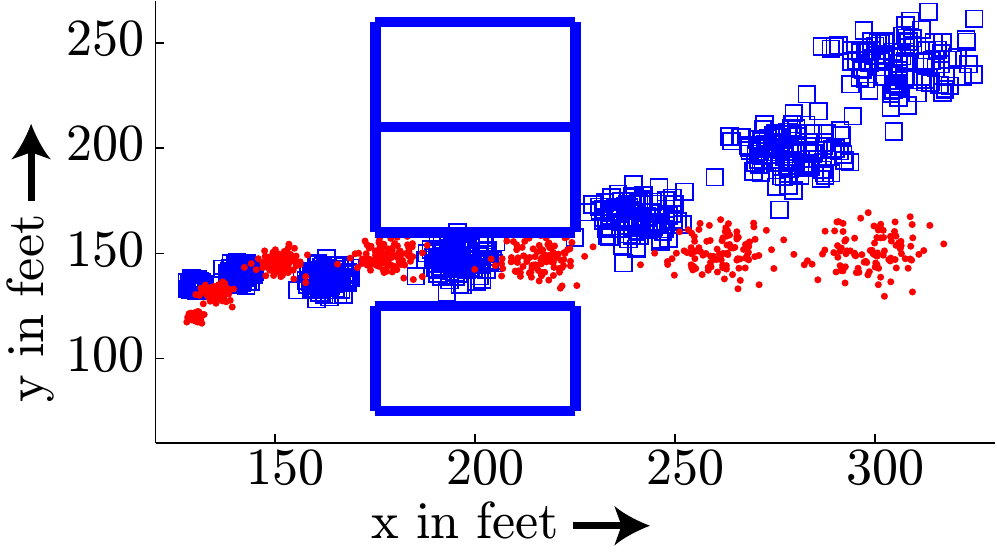}
  \caption{The plot shows the trajectories of two UAVs passing a gap between obstacles (depicted as solid blue boxes) computed with our novel RIPP constraints. 
    The lower UAV depicted by red dots waits for the upper UAV depicted by blue squares to pass the bottleneck first. 
%     The run-time for this example on a standard desktop PC was 100 \second. 
    The run-time for the same scenario computed with the sample-based approximation (SA) was almost 50 times longer than with our novel approach, while the optimal value of the control objective is only $4 \%$ better.}
  \label{fig:sample}
%   \vspace*{-2em}
\end{figure}

\subsection{Example Scenario}

In Figure~\ref{fig:sample}, we plot the trajectories of two UAVs, computed with the RIPP constraints. 
The planning horizon for this example is set to $H=7$, the starting parameters are $\vec{\mu}_1 = [130,135]^T$, $\vec{\mu}_2 = [130,120]^T$, $\vec{Z}_1 = [300, 250]^T$ and $\vec{Z}_2 = [300,150]^T$ with $N=100$ samples for each UAV's uncertain state. 
% Formulation DM takes 100 \second $\ $ to compute a solution, while formulation SA takes more than 67 \minute. 
The algorithm with our RIPP constraints solved the MILP to a value of the control objective of $ h_{min} = 146.8437$ while the optimal solution with sample-based approximation (SA) of inter-agent collision probabilities achieved only a slightly better value of $h_{min} =  141.2349$. 
The run-time was more than 40 times shorter with the RIPP constraints with only a small degree of sub-optimality in the value of the control objective. 

From the trajectories in Fig.~\ref{fig:sample} it can be seen that the lower UAV lets the upper UAV pass through the gap between the obstacles, since this behavior grants a better performance for both UAVs. 
Also it can be seen that our RIPP constraints do not force a strict separation of the samples of the uncertain states as some samples are allowed to fall below the minimum distance for example at the third time step. 
Our theoretical results on the RIPP constraints guarantee that the probability of a collision of the two UAVs does not exceed the predefined threshold.

\begin{figure}[htpb]
  \centering
  \includegraphics[scale = 1]{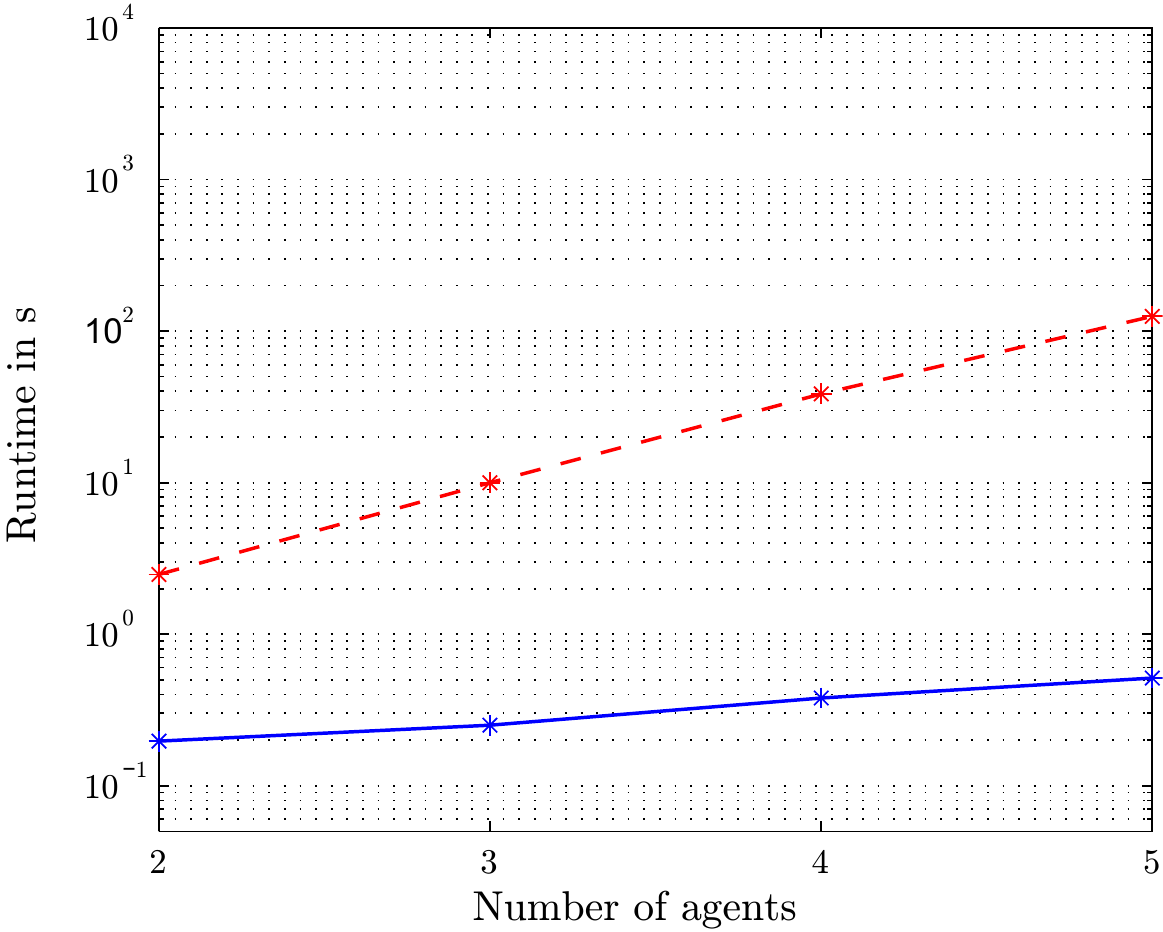}
  \caption{Depicted are the run-times of the controller with sample-based approximation of inter-agent collision probabilities (SA) in a red dashed line. The blue continuous line depicts the run-times of a controller with our novel RIPP constraints. 
%     The run-times are depicted for tests with two to five randomly placed UAVs. 
    It can be seen that the controller with our RIPP constraints has significantly lower run-times, making it feasible for applications with real-time requirements. 
    The controller with sample-base approximation of inter-agent collision probabilities has prohibitively high run-times even for small numbers of agents. 
    Please notice that the vertical axis is in $\log$-scale.  }
  \label{fig:runtime_compare}
\end{figure}

\begin{figure}[htpb]
  \centering
  \includegraphics[scale = 1]{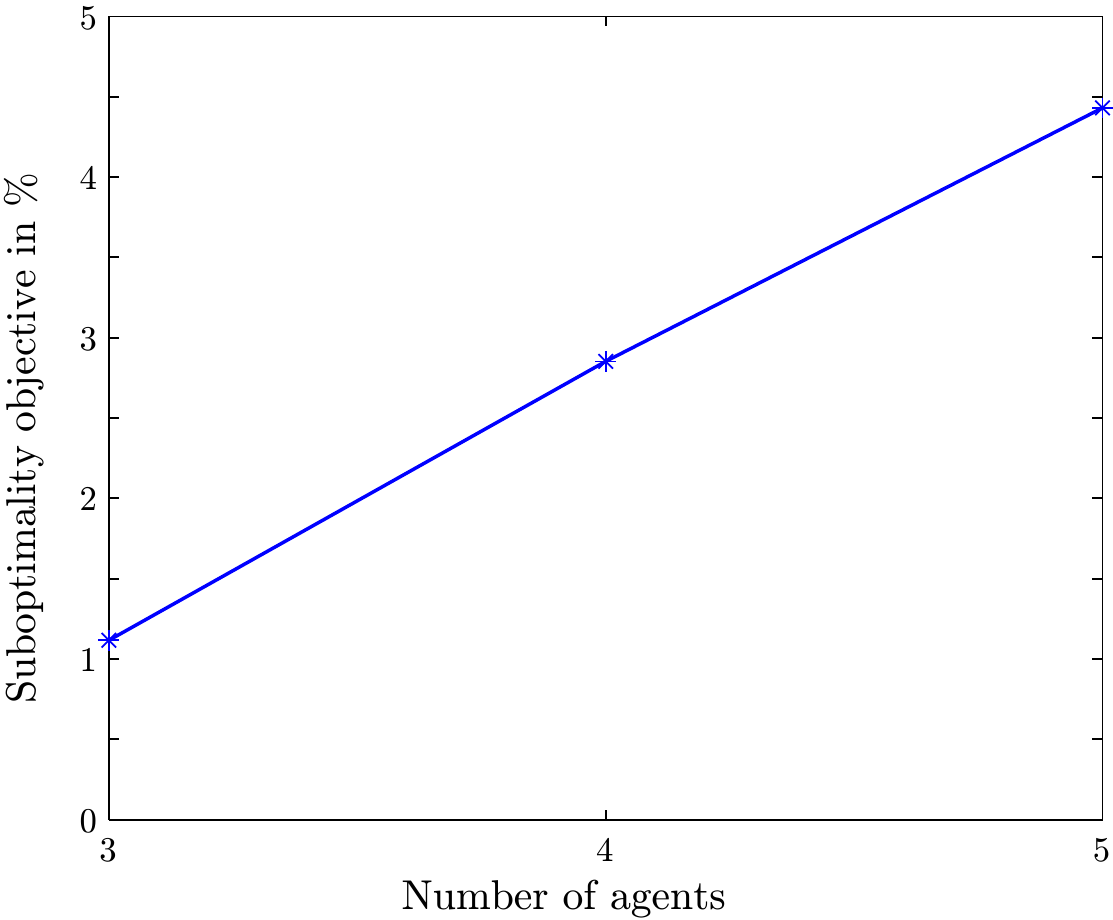}
  \caption{Depicted is the sub-optimality of controls found with our novel RIPP constraints in terms of percent of the optimal objective. 
  It can be seen that although much faster run-times are achieved the degree of sub-optimality is less than $5 \%$ and hence very low. 
  This plot is read in the way that for example for three agents the degree of sub-optimality in the optimal objective of the RIPP controller is a little above $1\%$. 
  We omitted the sub-optimality for two agents since the percentage was very low and would have not been noticeable on this plot. }
  \label{fig:suboptim}
\end{figure}

% \begin{figure}[h]
%   \centering
%   \includegraphics[scale = 1]{fig/25runs30particles_mb }
%   \caption{The blue continuous line depicts the run-times of a controller with our proposed conservative constraints (DM). The run-times are depicted for an increasing number of agents. These are the same results as in Fig.~\ref{fig:runtime_compare} only a bit more zoomed in. Please note that the experiments were set up randomly but in a way to guarantee that a maximum number of conflicts will occur, e.g. by guaranteeing that agents will have to pass each other. The run-times for up to four agent number sare very good. Interactions between five and more agents in a period as short as the planning horizon will most likely not occur. }
%   \label{fig:mb}
% \end{figure}
\subsection{Quantitative Results}

In Fig.~\ref{fig:runtime_compare}, we compare the run-time of a controller with the sample-based approximation of the collision probabilities (SA) with a controller with our novel RIPP constraints for collision avoidance. % the conservative constraints on the distance of the means (DM) from Sec.~\ref{sec:col_av_app}. 
The UAVs' starting positions and their target way points were randomly drawn to lie within a certain area to insure the occurrence of inter-agent collisions during planning. 
All results shown are averaged over 50 Monte-Carlo runs with each UAV's uncertain state approximated by 30 samples. 
The run-times of the MPC algorithm with RIPP constraints for the same Monte-Carlo runs can also be found in Table~\ref{table:table}. 
% The number of samples is so low because for a higher number of samples SA had a too high memory usage and did not run stable. 
% We compared the run-times of both approaches and the optimal objective of the 
It can be seen that the controller SA has prohibitively high run-times even for small numbers of agents. 
Our proposed approach based on the RIPP constraints on the other hand performs performs well with much lower run-times. 

\begin{table}
  \begin{tabular}{|p{3cm}||p{2cm}|p{2cm}|p{2cm}|}
    \hline 
    
				&	\multicolumn{3}{|c|}{Run-time RIPP constraints ($\second$)} \\ \hline
	Number of agents	&	$\min$				&	$\max$				&	mean 		\\ \hline
		      2 	&	0.14				& 	1.12				& 	0.19		\\ \hline
		      3		&	0.22				&	0.29				&	0.25		\\ \hline
		      4		&	0.28				&	1.09				&	0.37		\\ \hline
		      5		&	0.39				&	0.73				&	0.51		\\ \hline
  \end{tabular}
  \caption{Run-times of planning with RIPP constraints, same results as in Fig.~\ref{fig:runtime_compare}. }
  \label{table:table}
\end{table}

In Fig.~\ref{fig:suboptim} we plot the sub-optimality of our proposed approach based on RIPP constraints in percent of the optimal objective of the controller SA. 
It can be seen that although much faster run-times are achieved the degree of sub-optimality measured in the optimal value of the objective function is very low. 
This is a particularly important result as the controller with the sample-based approximation of inter-agent collision probabilities converges against the ``true'' controller with the ``true'' constraints on inter-agent collision probabilities for increasing sample numbers. 
Hence, it represents a benchmark, since its optimal solution is guaranteed to be close to the \emph{``true'' optimal solution}. 
This means that a controller with our RIPP constraints only introduces small sub-optimality compared to a controller that is close to the ``true'' optimal controller. 
The slight increase in sub-optimality for higher agent numbers stems from the fact, that the RIPP constraints introduce some conservatism and the degree of conservatism adds up for more agents in the system. 
% 
% The run-times depicted in Fig.~\ref{fig:mb} are the same results as in Fig.~\ref{fig:runtime_compare} only a bit more zoomed in. 
% Please note that all experiments were set up randomly but in a way to guarantee that a high number of conflicts will occur, e.g. by placing start and goal positions of agents randomly in a very confined space, guaranteeing that agents will have to pass each other during the planning horizon. 
% The run-times for up to four agent number are very good even under these extreme conditions. 
% Interactions between five and more agents in a period as short as the planning horizon $H=7$ will most likely not occur in real-world scenarios so that our proposed approach is expected to behave very well under real-world conditions. 
% All results in this section are averaged over 50 Monte 
% In average with RIPP constraints, a significantly lower run-time can be achieved, while the degree of sub-optimality measured in the value of the control objective of the solution is at a very low level. 
% Also, the deviation in run-time measured in the standard deviation of the run-times of the Monte-Carlo runs is
 %, hence the similar magnitudes for different numbers of agents in Table~\ref{table1}. 
\begin{figure}[htbp]
  \centering
  \subfigure[Robust coupling constraints.]{
  \includegraphics[scale = 0.5]{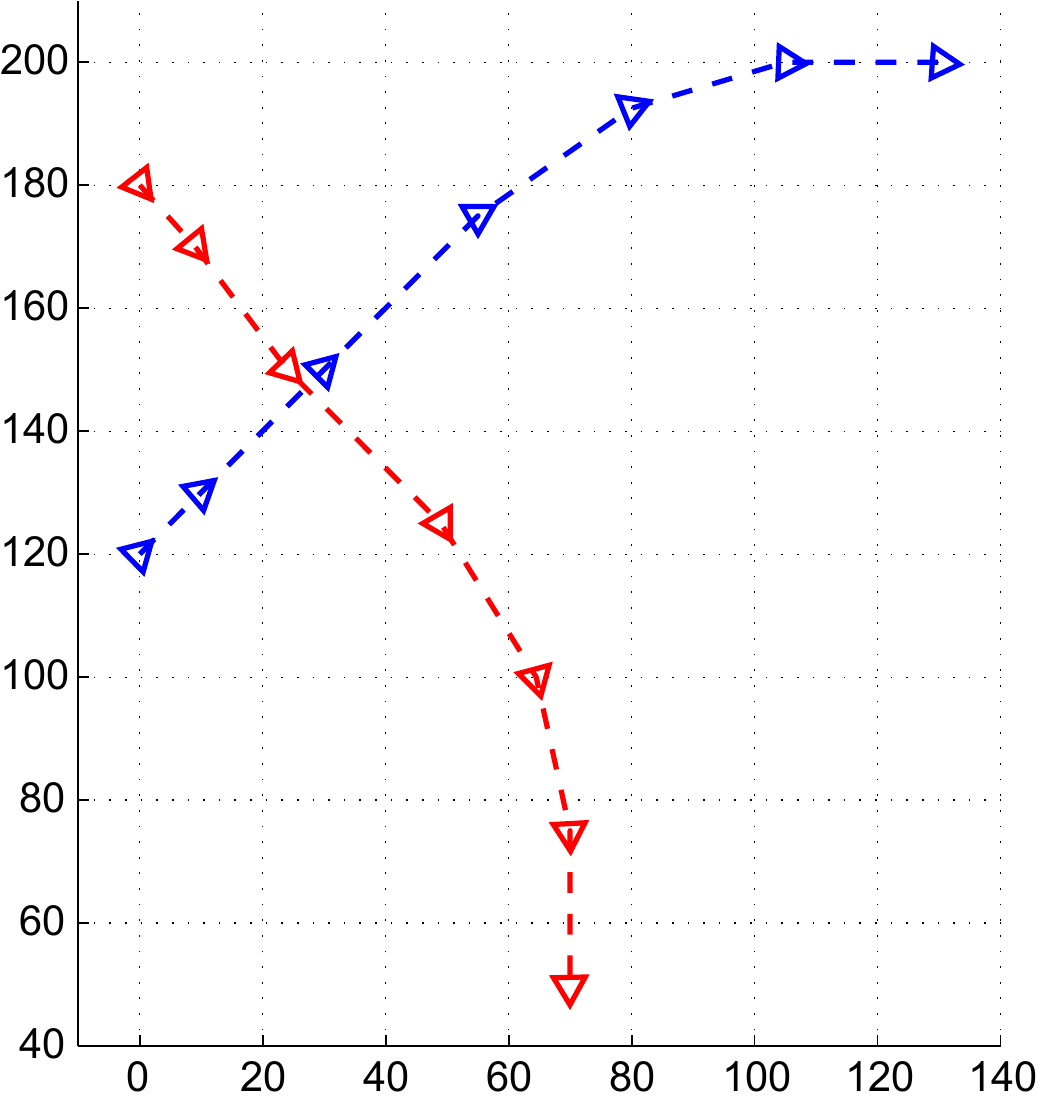}}
  \subfigure[Probabilistic coupling constraints.]{
  \includegraphics[scale = 0.5]{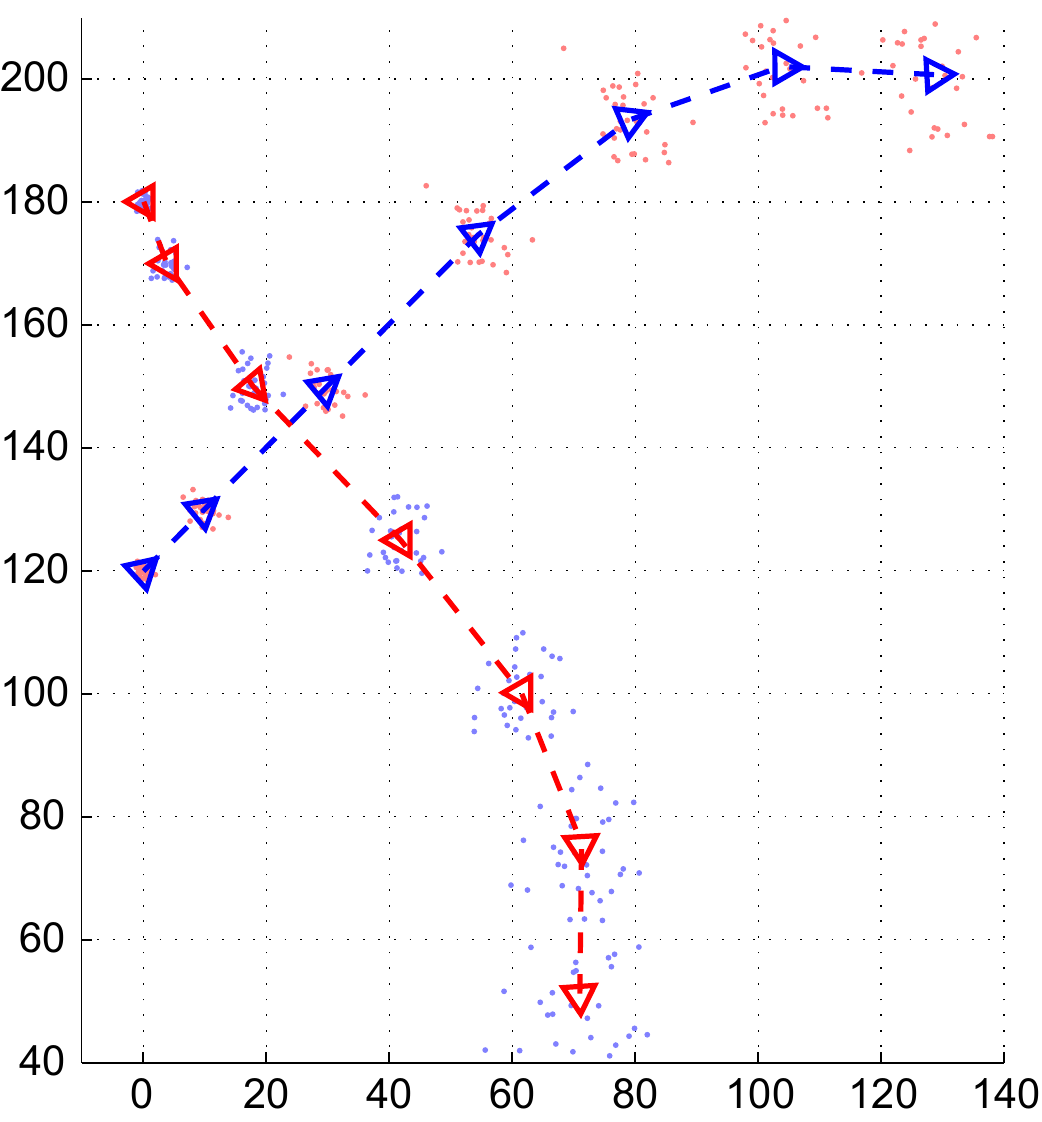}}
  \caption{Trajectory for two UAVs with robust and probabilistic coupling constraints for collision avoidance.
	  The robust contraints allow the UAVs to come closer in the third time step than the probabilistic constraints. 
	  The robust plan has a probability of a collision of more than $28 \%$ in the third time step, the probabilistic plan of $0.7\%$.
	  The chance constraint bound for collision probabilities was set to $1\%$.
	  It can be seen that the UAVs take much higher risks with the robust constraints, while the probabilistic constraints allow to precisely control how cautious the UAVs behave or how much risk they take. 
	   }
  \label{fig:robust_traj}
\end{figure}
\subsection{Comparison with Robust Control}
% \begin{itemize}
%   \item Gibt keine anderen Ans\"atze zu Multi-Agenten Regelung mit Constraints an inter-agent collisions Wahrscheinlichkeiten
%   \item Deshalb vergleichen wir uns mit einem Ansatz mit robusten Kupplungsconstraints f\"ur die Kollisionsvermeidung aus centralized robust MPC \cite{Richards2005}
%   \item Dieser Ansatz geht auch davon aus, dass die Regelung der Agenten durch exogene St\"orungen beeinflusst wird, modelliert diese aber nicht stochastisch, sondern unknown but confined to lie in a bounded set
%   \item Die Regelung soll dann robust constraint satisfaction garantieren dh. die Constraints sollen erf\"ullt sein f\"ur den gr\"osst m\"oglichen Ausschlag der St\"orungen.
%   \item Wir beschr\"anken und in unserem Vergleich nur auf Constraints, die die Interaktion von Agenten modellieren.
%   \item F\"ur diesen Vergleich verwenden wir das gleich System modell, stochastische St\"orungsmodell und Zielfunktion, die Schranken an Beschleunigung werden als $\Vert \vec{u}^i \Vert_\infty \le 10$ gew\"ahlt. 
%   \item Die f\"ur die robuste Regelung modellieren wir in \cite{Richards2005} die St\"orungen beschr\"ankt durch $a_{max}$ was $10 \%$ der Control force ist.
%   \item Ein nilpotenter Controller $\mat{K}$ f\"ur unser System $(\mat{A}, \mat{B})$ ist $\mat{K} = [-1 0 -1.5 0; 0 -1 0 -1.5]$
%   \item Die Zustandconstraints die sich daraus bestimmen lassen sind $\Vert \vec{x}^1_t - \vec{x}^2_t\Vert_\infty > 2a_{\max}$ f\"ur $t > 1$ f\"ur zwei Agenten $1$ und $2$. 
% \end{itemize}
To the best of our knowledge there are no other approaches to model chance constraints on the probability of inter-agent collisions, so we compare our approach to robust control approaches. 
We compare our approach to coupling constraints on the states of agents from the centralized robust MPC literature for the control of multiple UAVs~\cite{Richards2005}. 
This work also assume that the UAVs systems are affected by exogenous disturbances but does not model them stochastically but assumes them to be unknown but confined to lie in a bounded set. 
So the model of the UAVs dynamics is as in Eq.~\ref{eq:UAV_dynamics} without stochastic disturbance $\vec{\nu}^i_t$ but with unknown but bounded disturbance with $\Vert \vec{\nu}^i_t \Vert_{\infty} \le a_{\max}$. 
The aim of the MPC controller is to guarantee robust constraint satisfaction, i.e. constraints on the states of the UAVs have to be satisfied under the strongest possible disturbance. 

In our comparison we restrict ourselves to constraints that model interactions between agents. 
We use the same system dynamics as above, the same stochastic disturbance model for our approach, the same objective and the bounds on the control inputs are set to be bounded by $\Vert \vec{u}^i \Vert_\infty \le 10$. 
Like in~\cite{Richards2005} we model the disturbances for robust control to be bounded by up to $10 \%$ of the control input, hence, the disturbances acting on the states are bounded by $a_{\max} = 1$. 
The constraints on the positions of UAV $1$ and UAV $2$ resulting from the constraint tightening in~\cite{Richards2005} are
\begin{align}
 \Vert \vec{x}^1_t - \vec{x}^2_t \Vert_{\infty} > \epsilon + 2\alpha(t)
\end{align}
where
\begin{align}
 \alpha(1) = 0 \ ,\ \alpha(2) = 0 \ , \  \alpha(t) = \Vert [1\ 0\ 0\ 0] \mat{L}\mat{B} \Vert a_{\max} \text{ for } t>2 \ .
\end{align}
and $\mat{L} = \mat{A} + \mat{B}\mat{K}$. 
The matrix $\mat{K}$ is a two-step nilpotent controller $\mat{K}$ for the system $(\mat{A}, \mat{B})$ and it can be checked that
\begin{align}
 \mat{K} = \begin{bmatrix}
            -1 & 0 & -2 & 0 \\
	     0 & -1&    0 & -2
           \end{bmatrix}
\end{align}
fulfills the requirements that $\mat{L}^2 = \mat{0}$.

In Figure~\ref{fig:robust_traj} we plot the trajectories for two UAVs computed with robust constraints and the probabilistic constraints for collision avoidance. 
From the plot it becomes apparent that the robust contraints allow the UAVs to come closer in the third time step than the probabilistic constraints do.  
This fact is also reflected in the collision probabilities for this time step: the robust plan has a probability of a collision of more than $28 \%$ in the third time step, the probabilistic plan of $0.7\%$. 
These collision probabilities were determined with $1000000$ Monte-Carlo samples. 
% The robust plan has a probability of a collision of more than $28 \%$ in the third time step, the probabilistic plan of $0.7\%$. 
The chance constraint bound for collision probabilities was set to $\delta^{1,2}_t = 1\%$ at each time step $t$. 

It can be seen that the UAVs take much higher risks with the robust constraints, while the probabilistic constraints allow to precisely control how cautious the UAVs behave or how much risk they take. 
It could be possible to adjust the robust constraints in such a way that the UAVs behave more cautiously for example by raising the bound $a_{\max}$ on the unknown but bounded disturbances. 
But the question remains on how to adjust this bound such that a pre-specified collision probability can be guaranteed? 
The probabilistic constraints allow the user to precisely specifiy upper bounds on these collision probabilities and, hence, also the probability of a failure of the UAV mission while with robust constraints it is not clear how this could be done.

% \begin{itemize}
%   \item Conducted 
% \end{itemize}
% \begin{figure}[h]
%   \centering
%   \includegraphics[scale = 1]{fig/2agents_3obstacles_sample_nice_ai.pdf}
%   \caption{The plot shows the trajectories of two UAVs computed with the conservative approximation with constraints on the distance of the means of the agents (DM). 
%     The lower UAV depicted by red dots waits for the upper UAV depicted by blue squares to pass the bottleneck first. 
%     The runtime for this example on a standard desktop PC was 100 \second. The optimal value of the control objective it achieved was 146.8437. The runtime for the same scenario computed with the sample approximation (CC) took more than 67 \minute. The value of the control objective was 141.2349.}
%   \label{fig:sample}
% %   \vspace*{-2em}
% \end{figure}
\section{Conclusions}\label{sec:concl}
In this work we gave two methods to formulate chance constraints on the probability of inter-agent collisions in order to make them tractable for a mixed integer linear optimization routine. 
% The first is a direct approximation of the coupling constraints through sample approximations. 
The first formulation is a straightforward sample-based approximation of the probability of collisions between agents with convergence against the true probability as the number of samples goes to infinity. 
The number of binary variables this formulation introduces to the optimization problem, however, depends quadratically on the number of samples in the approximation of the agents' state distributions. 
This renders the approach computationally infeasible since the quality of the sample approximation of the chance constraints improves with the number of samples. 

To overcome such limitations, we introduced alternative collision avoidance constraints that couple the agents' control problems for coordination but are independent of the number of samples. 
We constructed a region of increased probability of presence (RIPP) for the uncertain positions of the agents and introduced constraints that these RIPP regions do not overlap for differing agents. 
Because the construction of the RIPP regions is based on a probabilistic inequality we were able to prove that controls that satisfy the RIPP constraints are automatically feasible for MPC problem with chance constraints on inter-agent collision probabilities. 

It is remarkable that the RIPP constraints can guarantee this, solely based on means and covariances of the uncertain states of the agents without the need to evaluate the complicated inter-agent collision probabilities. 
Also, since the probabilistic inequality we use to determine the RIPP regions holds for uncertain states with arbitrary state distribution, our approach is not limited to Gaussian state distributions but is generally applicable. 
Further we demonstrated in our simulations that the RIPP constraints not only lead to theoretically feasible controls but also that the sub-optimality of these controls is very low compared to controls computed with the almost optimal controls computed with the sample-based approximation of collision probabilities. 
Compared to robust collision avoidance constraints we showed in a simulation that our probabilistic constraints are better suited for situations in which the systems under control are affected by stochastic disturbance. 
% These alternative constraints are constraints on the distance of the sample means of the state estimates of the agents and conservatively bound the probability of inter-agent collisions. 
% It has the advantage that it holds for general multivariate estimates of the agents' states, such as ones arising from nonlinear filters. 
% The number of binary variables in the resulting MILP is completely independent from the number of samples in the approximation of the state distributions. 

% % \begin{itemize}
% %   \item it holds for general multivariate estimates of the agents' states, such as ones arising from nonlinear filters and
% %   \item the complexity of the resulting MILP, measured in the number of binary variables, does not depend on the complexity of the sample representation of the state estimates.
% %   \item it generates plans that are feasible to the 
% %   \item it automatically adapts itself to the quality of the state estimate and
% %   \item it explicitly accounts for correlations in the agents' state estimates.
% % \end{itemize}
Distributed control strategies are advantageous in multi-agent systems, so future work will be concerned with applying distributed MILP solving techniques to achieve a decentralized architecture \cite{Vanderbeck2000,Hong2011}.

\section*{Acknowledgements}
This work was partially supported by the German Research Foundation (DFG) within the Research Training Group GRK~1194 ``Self-organizing Sensor-Actuator-Networks''.
Jan Calliess is grateful for funds via the UK EPSRC "Orchid" project EP/I011587/1.

\bibliographystyle{IEEEtran}
\bibliography{literature.bib}

\end{document}